\documentclass[10pt]{article}

\usepackage{amsmath}
\usepackage{amsfonts}
\usepackage{amssymb}
\usepackage[numbers]{natbib}
\usepackage{geometry}
\usepackage{graphicx}

\usepackage{bbm}
\usepackage{amsthm}
\theoremstyle{definition} 

\theoremstyle{theorem} 

\usepackage{hyperref}
\hypersetup{
   colorlinks   =  true,
    linkcolor    = cyan,
    citecolor    = red,
     urlcolor	=magenta,     
}

\DeclareMathOperator\spn{span}

\begin{document}

\hfill\today

\begin{center}
\ 

\bigskip

{\huge
{Fuzzy worldlines with $\kappa$-Poincar\'e symmetries}}
\bigskip
\bigskip
\end{center}

\begin{center}

{\sc Angel Ballesteros$^1$, Giulia Gubitosi$^{3,4}$, Ivan Gutierrez-Sagredo$^{1,2}$,  Flavio Mercati$^{1}$}

\medskip

{$^1$Departamento de F\'isica, Universidad de Burgos, 
09001 Burgos, Spain}

{$^2$Departamento de Matem\'aticas y Computaci\'on, Universidad de Burgos, 
09001 Burgos, Spain}

{$^{3}$ Dipartimento di Fisica ``Ettore Pancini'',
Universit\`a di Napoli Federico II, Napoli, Italy}

{$^{4}$ INFN, Sezione di Napoli}

 \medskip
 
e-mail: {\href{mailto:angelb@ubu.es}{angelb@ubu.es}, \href{mailto:giulia.gubitosi@unina.it}{giulia.gubitosi@unina.it}, \href{mailto:igsagredo@ubu.es}{igsagredo@ubu.es}, \href{mailto:fmercati@ubu.es}{fmercati@ubu.es}}

\end{center}

\medskip

\begin{abstract}

A novel approach to study the properties of models with quantum-deformed relativistic symmetries relies on a noncommutative space of worldlines rather than the usual noncommutative spacetime. In this setting, spacetime can be reconstructed as the set of events, that are identified as the crossing of different worldlines. We lay down the basis for this construction for the $\kappa$-Poincar\'e model, analyzing the fuzzy properties of $\kappa$-deformed time-like worldlines and the resulting fuzziness of the reconstructed events.

\end{abstract}

\bigskip



\tableofcontents

\section{Introduction}

Many approaches to the quantum gravity problem focus on defining the new properties that spacetime should acquire at the Planck scale, where it is expected that a description of spacetime in terms of a smooth pseudo-Riemannian manifold is no longer available \cite{Mead:1964zz, Amati:1988tn, Ahluwalia:1993dd, Garay1995}.  The new formalisms that have been developed mostly attempt to introduce some sort of discretization  (see  \cite{Loll:2019rdj, Surya:2019ndm} and references therein) or noncommutativity \cite{Snyder1947, DFR1994, MW1998, Szabo2003, FL2006, Amelino-Camelia:2008fcv, Balachandran:2010gc}.

Here, we take a different approach, that is closer to the spirit of Einstein's operative construction of spacetime based on timing the emission and reception of light signals. From this perspective, spacetime is just a collection of interaction events, defined by the  crossings of  worldlines. Clearly, in a classical spacetime these crossings define sharp points. Once quantum gravity effects are taken into account this might not be the case anymore. For example,  close to the Planck scale  a deformation of relativistic symmetries might emerge, such that the Planck scale can be accommodated as a relativistic invariant scale \cite{Amelino-Camelia2001testable,Kowalski-Glikman2001, Amelino-Camelia2002planckian,  MS2002,LN2003versus,BRH2003newdoubly, FKS2004gravity, ASS2004,Amelino-Camelia2010symmetry}. In this case, particle interactions might appear nonlocal to a far-away observer, while retaining their locality from the point of view of local observers. This led to the development of relative-locality theories \cite{AFKS2011deepening, AFKS2011principle, AAKRG2012relativelocality}.

Our work lies in the same general framework as relative locality, since our  setup is given by deformed relativistic symmetries. In relative locality, these are used to work out the properties of the momentum space, which turns out to be curved, with the Planck scale setting the scale of curvature \cite{Kowalski-Glikman2013living, GM2013relativekappa, Amelino-Camelia:2013sba, BGGH2017curvedplb, BGGH2018cms31}. Then the propagation and interaction of particles is described in terms of a Lagrangian or Hamiltonian theory built on momentum space as base manifold.
Recently, some of us pointed out \cite{BGH2019worldlinesplb} (see also~\cite{BRH2017}) that (deformed) relativistic symmetries naturally endow the space of worldlines -- namely, the space of all possible trajectories of particles --  with an additional noncommutative structure which can be used to construct a  picture that is complementary to that of relative locality. In this setting, the Planck scale parameter  governs the noncommutativity of the `quantum' space of worldlines.

In special relativity, the space of worldlines can be obtained as the homogeneous space of the Poincar\'e transformations modulo rotations and time translations, that are the transformations leaving invariant the origin of this space (represented by a worldline corresponding to  a particle that is stationary at the origin of the coordinates system). This is completely analogous to the construction of spacetime as the homogeneous space of the Poincar\'e transformations modulo the Lorentz subgroup, which leaves the origin invariant. If one takes a quantum group as the group of deformed relativistic symmetries \cite{Majid1988, ChariPressley1994, Majid1995Book}, then the associated spacetime turns out to be noncommutative \cite{LRNT1991,LNR1992fieldtheory,MR1994,BHOS1995nullplane,Zakrzewski1997, BRH2003minkowskian,BLT2016unified, BLT2016unifiedaddendum, MS2018constraints, BM2018extended, Juric:2015jxa}.
Similarly, the resulting space  of worldlines will inherit the quantum properties of the symmetry group, and so will  spacetime events, defined as worldlines crossings. In this work we are going to characterize the properties of such noncommutative worldlines and their interactions. Following \cite{BGH2019worldlinesplb}, we take the much-studied $\kappa$-Poincar\'e group \cite{LRNT1991,LNR1992fieldtheory,MR1994, Meljanac:2012fa}  as the deformed group of symmetries. 

Because in our setting worldlines are not sharply defined, stating whether they cross is a nontrivial task. To this aim, we define an \emph{impact parameter}, which classically quantifies the minimum spatial distance achieved by two worldlines at equal times. This parameter is zero when two classical worldlines cross (thus defining an event) and indicates how far they pass from each other otherwise. When worldlines are fuzzy, so will be this parameter, and one can study its expected value and variance. In particular, this parameter can be used to identify a fuzzy region of spacetime where two worldlines are likely to cross, which corresponds to a fuzzy event. 

We find that, as expected based on the results of relative locality, the larger the expectation value of the impact parameter, the larger its uncertainty. We study this feature quantitatively, and are able to find two different regimes, one corresponding to sharp values of the worldlines velocity and large variance of the intercept, and the other one corresponding to large variance in both the velocity and the intercept. In the first case the uncertainty of the impact parameter grows linearly with the square root of its expectation value. This case is the one in closer relation to the relative locality framework, since it deals with massive particles with sharply defined momenta. In the second case the uncertainty grows linearly with the expectation value. For this second regime we do not find a correspondence in terms of the relative locality framework. However, we show that the two regimes correspond to analogous regimes found in studies of $\kappa$-Minkowski spacetime, one where the variance of the spatial coordinates is proportional in the variance of the time coordinate and one where the variance of the time coordinate is fixed, respectively.

The plan of this paper is the following. In section \ref{sec:classicalworldlines} we revisit the relation between worldline parameters and spacetime coordinates, and define the impact parameter, all in a classical setting. In section \ref{sec:kappadeformation} we revisit the results of  \cite{BGH2019worldlinesplb}, where  the space of $\kappa$-worldlines associated to the $\kappa$-Poincar\'e group of symmetries was first constructed. Section \ref{sec:quantumworldlines} is devoted to the characterization of the properties of the  $\kappa$-worldlines in terms of Hilbert space and quantum states built in it. We show that the origin in the space of worldlines admits a perfectly localized state, so it can be used as a sharp reference to compute the impact parameter of other (quantum) worldlines with respect to it.
In section \ref{Sec:FuzzyWorldlines} we quantitatively study the properties of the $\kappa$-worldlines and their interactions. In order to make the analysis feasible, we resort to a semiclassical approximation, taking the classical probability distributions  corresponding to the Wigner quasiprobability of a  quantum wavefunction. This provides us with results that accurately reproduce the statistical properties of quantum states.


\section{Classical worldlines in Minkowski spacetime}
\label{sec:classicalworldlines}

We start by reviewing the classical description of worldlines in special relativity, which allows us to establish a correspondence between Minkowski spacetime coordinates $x^{\mu}$ and the parameters which identify a worldline, namely its velocity $\mathbf{v}$ and intercept at time zero, $\mathbf{B}$.  Using these, we are able to define an `impact parameter' between two worldlines, which is zero when the two worldlines cross. This defines an event, that is a point in spacetime.

Let us consider Minkowski spacetime, i.e the four-dimensional vector space with the pseudo-Riemannian metric 
\begin{equation}
ds^2= (d x^0)^2 - (d x^1)^2 -( d x^2)^2 - (d x^3)^2 \, ,
\end{equation}
and introduce linear coordinates $x^\mu$ on it.\footnote{
Latin indices run from $1$ to $3$, denoting  3-vectors by $\mathbf{v}=(v^1,v^2,v^3)$, whereas greek indices run from $0$ to $3$, denoting 4-vectors by $v^{\mu} =(v^0,v^1,v^2,v^3)$. We use units such that the speed of light $c=1$. } As it is well-known, the Levi-Civita connection for Minkowski spacetime is flat, so  the geodesic equation is just $\frac{d^2 x^\mu (\tau)}{d \tau^2} = 0$, with $\tau$ the proper time. Integration is straightforward and leads to 
\begin{equation}
\label{eq:wl_lineara}
x^\mu (\tau) = a^\mu \tau + b^\mu,
\end{equation}
with $a^{\mu}$ and $b^{\mu}$ integration constants.
Using $\tau = \frac{x^0-b^0}{a^0}$ one can write the equation of a geodesic as
\begin{equation}
x^i=\frac{a^i}{a^0}x^0 +(b^i-b^0\frac{a^i}{a^0})\equiv \frac{a^i}{a^0} x^0 +B^i \, .
\end{equation}
By means of this parameterization we can identify the coordinate velocity   $v^i\equiv  \frac{d x^i}{d x^0}=\frac{a^i}{a^0} $, such that 
\begin{equation}
\label{eq:wl_linearv}
x^i = v^i x^0 + B^i .
\end{equation}
This parameterization of worldlines establishes a map between spacetime coordinates $x^{\mu}$ and worldline parameters $(v^{i},B^{i})$. 

From now on, for reasons that will be made clear in the following section, we only consider time-like geodesics, so $\|\mathbf{v}\| < 1$. Then each worldline represents  a free massive particle that passes through the spacetime point $(0, \mathbf{B})$ with spatial velocity $\mathbf{v}$. In particular, this interpretation allows us to describe an especially simple worldline, which we call $w_{0}$, corresponding to a particle staying at the origin with zero velocity. This worldline will be taken to be the origin of the space of worldlines and is described by the equations $x^i (x^0)= 0$ or, equivalently, 
\begin{equation}
\label{eq:wl_0}
x^0 (\tau) = \tau, \qquad x^i (\tau)= 0.
\end{equation}

\subsection{Impact parameter of worldlines and definition of events}
\label{sec:impact}

As we mentioned, we aim at characterizing spacetime as a collection of events, defined as the crossing of (at least) two worldlines. Classically, it is straightforward to establish whether two worldlines cross, but this is not the case if they are somehow fuzzy, as it might be the case once quantum gravitational effects are taken into account. For this reason, we introduce a quantity that allows us to characterize the  distance between two worldlines. We call this quantity the \emph{impact parameter} $\beta(w_1,w_2)$ between two worldlines $w_1$ and $w_2$. It is defined  as the infimum of the squared spatial distance between the worldlines when they have the same coordinate time $x^0$:
\begin{equation}
\label{eq:impact}
 \beta(w_1,w_2) = \inf_{x^0\in \mathbb{R}} \sum_{i=1}^{3} (x^i_{(1)} - x^i_{(2)})^2.
\end{equation}
In the classical case, the impact parameter is zero when two worldlines cross, and the event happens at the time $x^{0}$ where  $\sum_{i=1}^{3} (x^i_{(1)} - x^i_{(2)})^2=0$. 
Eq.~(\ref{eq:impact}) is not Lorentz invariant in general, but the statement that $\beta=0$ is a Lorentz-invariant one, which expresses the fact that the two worldlines cross somewhere. 

If we consider fuzzy worldlines $\beta$ cannot vanish sharply, so the issue of finding a  Lorentz invariant analogue of \eqref{eq:impact} becomes more pressing.  A possible solution is to  look for the extremum of $\beta$ over all reference frames.  In the classical case it can be shown that  this extremum  is found in the reference frame where one of the worldlines is $w_{0}$,  given by \eqref{eq:wl_0}. Such  proof is  tricky to repeat in the fuzzy case, because it would require us to compute explicitly the action of deformed Lorentz transformations on two generic worldlines. So in this paper we will not study the properties of $\beta$ for generic fuzzy worldlines $w_{1}$ and $w_{2}$, but we will focus on the case where one of the two worldlines is $w_{0}$ \footnote{Clearly, if $w_{0}$ is a fuzzy worldline, the issue with Lorentz invariance is not  completely solved. However, in section \ref{perfecly_localized_state} we will demonstrate that this reference worldline $w_{0}$ can be chosen sharply.}.
In this case the impact parameter reads 
\begin{equation}
\label{eq:distw0}
\beta(w_0,w) = \min_{x^0 \in \mathbb{R}} \sum_{i=1}^{3} (x^i)^2 = \min_{x^0 \in \mathbb{R}} \sum_{i=1}^{3} (v^i x^0 + B^i)^2 
\, ,
\end{equation}
and one can easily find a more explicit expression for $\beta(w_0,w) $. In fact, the function  $ \sum_{i=1}^{3} (v^i x^0 + B^i)^2$ has  one (global) minimum at 
\begin{equation}
x^0_{min} = -\frac{\mathbf{v} \cdot \mathbf{B}}{\|\mathbf{v}\|^2}.
\end{equation}
Putting this back into \eqref{eq:distw0}, and using $\|\mathbf{v} \times \mathbf{B}\|^2 = \|\mathbf{v}\|^2 \|\mathbf{B}\|^2 - (\mathbf{v} \cdot \mathbf{B})^2$, one finds 
\begin{equation}\label{eq:distw0explicit}
\begin{array}{ll}
\beta(w_0,w) = \frac{\|\mathbf{v} \times \mathbf{B}\|^2}{
\|\mathbf{v}\|^2} &\quad\text{if} \quad \mathbf{v}  \neq 0\\
\\
\beta(w_0,w) =  \|\mathbf{B}\|^2 & \quad\text{if} \quad \mathbf{v} = 0.
\end{array}
\end{equation}
 In terms of this expression for the impact parameter, the occurrence of a classical event admits a clear interpretation: the event only takes place if the position and velocity vectors are parallel (if the velocity of the second worldline is zero, then there is no event, since the two worldlines describe particles that are at rest at different spatial points).

Notice that the impact parameter function \eqref{eq:distw0explicit} can be generalized to find the explicit expression of the impact parameter  of two generic worldlines, eq.~\eqref{eq:impact}:
\begin{equation}\label{eq:dist}
\beta(w_1,w_2)= \frac{\|\Delta \mathbf{v}\times\Delta\mathbf{B}\|^2}{\|\Delta\mathbf{v}\|^2}\,,
\end{equation}
with $\Delta \mathbf{v}=\mathbf{v_{(1)}}-\mathbf{v_{(2)}}$ and $\Delta\mathbf{B}=\mathbf{B}_{(1)}-\mathbf{B}_{(2)}$.  However, as we explained, this expression will not be needed in the following, since we will focus on the case described by  \eqref{eq:distw0explicit}.


\section{$\kappa$-Minkowski spacetime and the $\kappa$-space of worldlines}
\label{sec:kappadeformation}

In this section we briefly review the construction of $\kappa$-Minkowski spacetime and of the $\kappa$-space of worldlines as homogeneous spaces of $\kappa$-Poincar\'e, following \cite{BGH2019worldlinesplb}. Before this,  we first look at the classical construction, in order to find the relation between the parameters $({\bf v},{\bf B})$ defined in the previous section, that are used to parameterize a worldline in  spacetime,  and the parameters $( y^\mu, {\eta^\mu})$ used to identify a worldline in the homogeneous space of worldlines. The same relation holds  in the $\kappa$-deformed case, which only differs from the classical case because of  the existence of a nontrivial Poisson structure induced by the $\kappa$-deformation, which will be later  transformed into a noncommutative algebra of worldline coordinates.

\subsection{Minkowski spacetime and its space of time-like worldlines}

Let us consider the Poincar\'e Lie algebra $\mathfrak{g}=\mathfrak{p}(3+1) \equiv \mathfrak{so}(3,1) \ltimes \mathbb{R}^4$, which generates the (3+1) Poincar\'e group $G = P(3+1)$. We will make use of the kinematical basis $\{P_0,P_a, K_a, J_a\}$  consisting of the generators of time translation, space translations, boosts and rotations, respectively. In this basis the commutation rules for $\mathfrak{p}(3+1)$ read
\begin{equation}
\begin{array}{lll}
[J_a,J_b]=\epsilon_{abc}J_c ,& \quad [J_a,P_b]=\epsilon_{abc}P_c , &\quad
[J_a,K_b]=\epsilon_{abc}K_c , \\[2pt]
\displaystyle{
  [K_a,P_0]=P_a  } , &\quad\displaystyle{[K_a,P_b]=\delta_{ab} P_0} ,    &\quad\displaystyle{[K_a,K_b]=-\epsilon_{abc} J_c} , 
\\[2pt][P_0,P_a]=0 , &\quad   [P_a,P_b]=0 , &\quad[P_0,J_a]=0  ,
\end{array}
\label{eq:ads_Liealg3+1}
\end{equation}
where sum over repeated indices is assumed.

Let us now consider the Lie subalgebras of $\mathfrak{g}$ given by 
\begin{equation}
\mathfrak{l} = \spn\{ K_a, J_a \}, \qquad \mathfrak{h} = \spn\{ P_0, J_a \},
\label{sub}
\end{equation}
corresponding to the Lie subgroups $L$ and $H$ of $G$, respectively. Geometrically, these Lie subgroups are just  the stabilizers of the origin of Minkowski spacetime  and  of the origin of the space of worldlines (time-like oriented geodesics), respectively.

 In order to describe the spacetime $\mathcal{M}$ and  the space of worldlines $\mathcal{W}$ as the coset spaces  $\mathcal{M}=G/L$ and $\mathcal{W}=G/H$, respectively, one needs to  exponentiate a faithful representation of the Lie algebra $\mathfrak{p}(3+1)$ in two different and carefully chosen orders. Explicitly, let us take  as faithful representation $\rho : \mathfrak{p}(3+1)\rightarrow \text{End}(\mathbb R ^5)$ such that a generic element $X$ of the Lie algebra $\mathfrak{p}(3+1)$ is given by
\begin{equation}
\label{eq:repG}
\rho(X)=   x^\alpha \rho(P_\alpha)  +  \xi^a \rho(K_a) +  \theta^a \rho(J_a) =\left(\begin{array}{ccccc}
0&0&0&0&0\cr 
x^0 &0&\xi^1&\xi^2&\xi^3\cr 
x^1 &\xi^1&0&-\theta^3&\theta^2\cr 
x^2 &\xi^2&\theta^3&0&-\theta^1\cr 
x^3 &\xi^3&-\theta^2&\theta^1&0
\end{array}\right) .
\end{equation}

In this framework, the  Minkowski spacetime $\mathcal{M}$ can be constructed as the homogeneous space given by the coset space  $\mathcal{M}=G/L$, where a generic element of the Poincar\'e group $G=P(3+1)$ is parametrized in the form
\begin{align}
\begin{split}
\label{eq:Gm}
&g_\mathcal{M}= \exp\{x^0 \rho(P_0)\} \exp\{x^1 \rho(P_1)\} \exp\{x^2 \rho(P_2)\} \exp\{x^3 \rho(P_3)\} \\
&\qquad\quad\times \exp\{\xi^1 \rho(K_1)\} \exp\{\xi^2 \rho(K_2)\} \exp\{\xi^3 \rho(K_3)\}
 \exp\{\theta^1 \rho(J_1)\} \exp\{\theta^2 \rho(J_2)\} \exp\{\theta^3 \rho(J_3)\} \, ,
\end{split}
\end{align}
and elements of the Lorentz subgroup $L$ are parametrized by
\begin{align}
\begin{split}
\label{eq:Lm}
&l= \exp\{\xi^1 \rho(K_1)\} \exp\{\xi^2 \rho(K_2)\} \exp\{\xi^3 \rho(K_3)\}
 \exp\{\theta^1 \rho(J_1)\} \exp\{\theta^2 \rho(J_2)\} \exp\{\theta^3 \rho(J_3)\}  .
\end{split}
\end{align}
In this way $x^\alpha$ are well-defined coordinates on the Minkowski spacetime $\mathcal{M}=G/L$.

In order to construct the space of worldlines as the homogeneous space $\mathcal{W}=G/H$ in a similar manner, new  local coordinates $(\eta^i,y^i)$ on the Poincar\'e group have to be introduced, in such a way that they give rise to well-defined coordinates on $\mathcal{W}$. To this aim, the Poincar\'e group $G=P(3+1)$ needs to be parametrized as follows:
\begin{align}
\begin{split}
\label{eq:Gw}
&g_\mathcal{W}=\exp\{\eta^1 \rho(K_1)\} \exp\{y^1 \rho(P_1)\}  \exp\{\eta^2 \rho(K_2)\}\exp\{y^2 \rho(P_2)\}  \exp\{\eta^3 \rho(K_3)\} \exp\{y^3 \rho(P_3)\} \\
&\qquad\quad\times
 \exp\{\phi^1 \rho(J_1)\} \exp\{\phi^2 \rho(J_2)\} \exp\{\phi^3 \rho(J_3)\} \exp\{y^0 \rho(P_0)\} \, ,
\end{split}
\end{align}
since in this way the isotropy subgroup of worldlines, parametrized as
\begin{align}
\begin{split}
\label{eq:Hw}
&h=  \exp\{\phi^1 \rho(J_1)\} \exp\{\phi^2 \rho(J_2)\} \exp\{\phi^3 \rho(J_3)\} \exp\{y^0 \rho(P_0)\},
\end{split}
\end{align}
is located at the right within the generic group element $g_\mathcal{W}$ and this guarantees that $(\eta^i,y^i)$ are well-defined coordinates on the coset space $\mathcal{W}=G/H$.

In order to find the relation between the two sets of coordinates,  the explicit matrix forms of the two group elements $g_{\mathcal M}$ and $g_{\mathcal W}$ has to be compared, thus finding (see~\cite{BGH2019worldlinesplb} for details)
\begin{equation}\label{eq:MWrelation}
x^\alpha= f^\alpha(y^\beta,\eta^i),\quad \xi^\alpha=\eta^\alpha,\quad \theta^\alpha=\phi^\alpha\,,
\end{equation}
where $f^\alpha$ are the following functions of the worldline coordinates $(y^\beta, \eta^i)$:
\begin{eqnarray}
\label{eq:f}
f^0(y^\mu,\eta^\mu)&=&y^1\sinh \eta^1+ y^2 \sinh\eta^2\cosh\eta^1+y^3\sinh\eta^3\cosh\eta^2\cosh\eta^1+y^0\cosh\eta^3\cosh\eta^2\cosh\eta^1\nonumber\\
f^1(y^\mu,\eta^\mu)&=&y^1\cosh \eta^1+ y^2 \sinh\eta^2\sinh\eta^1+y^3\sinh\eta^3\cosh\eta^2\sinh\eta^1+y^0\cosh\eta^3\cosh\eta^2\sinh\eta^1\nonumber\\
f^2(y^\mu,\eta^\mu)&=&y^2 \cosh\eta^2+y^3\sinh\eta^3\sinh\eta^2+y^0\cosh\eta^3\sinh\eta^2\nonumber\\
f^3(y^\mu,\eta^\mu)&=&y^3\cosh\eta^3+y^0\sinh\eta^3\,.
\end{eqnarray}

In the space of worldlines each point $(y^i,\eta^i)$ identifies one worldline $w$. In order to find the relation between these coordinates and the worldline parameters   $(B^i,v^i)$ defined in the previous section, we use the first of the equations above to write $y^0$ as a function of $x^0=f^0$: 
\begin{equation}
y^0=\frac{x^0}{\cosh\eta^3\cosh\eta^2\cosh\eta^1}- y^1\frac{\tanh \eta^1}{\cosh\eta^3\cosh\eta^2}- y^2 \frac{\tanh\eta^2}{\cosh\eta^3}-y^3\tanh\eta^3\,.
\end{equation}
Upon substituting this expression into the remaining equations in \eqref{eq:f} we get
 \begin{equation}
 \begin{array}{lcl}
f^1(y^\mu,\eta^\mu)&=&\frac{y^1}{\cosh \eta^1}+x^0\tanh\eta^1\\
f^2(y^\mu,\eta^\mu)&=&\frac{y^2}{ \cosh\eta^2}- y^1 \tanh \eta^1\tanh \eta^2+x^0\frac{\tanh\eta^2}{\cosh\eta^1}\\
f^3(y^\mu,\eta^\mu)&=&\frac{y^3}{ \cosh\eta^3}- y^2\tanh\eta^2\tanh\eta^3- y^1\frac{\tanh \eta^1\tanh \eta^3}{\cosh\eta^2}+x^0\frac{\tanh\eta^3}{\cosh\eta^2\cosh\eta^1}\,.
\end{array}
\end{equation}
Comparing these expressions with \eqref{eq:wl_linearv}, the following correspondence is found: 
 \begin{equation}
 \begin{array}{lcl}
 \label{eq:vi}
v^1(\eta^\mu)&=&\tanh\eta^1\\
v^2(\eta^\mu)&=&\frac{\tanh\eta^2}{\cosh\eta^1}\\
v^3(\eta^\mu)&=&\frac{\tanh\eta^3}{\cosh\eta^2\cosh\eta^1}\,,
\end{array}
\end{equation}
together with
\begin{equation}
 \begin{array}{lcl}
  \label{eq:Bi}
B^1(y^\mu,\eta^\mu)&=&\frac{y^1}{\cosh \eta^1}\\
B^2(y^\mu,\eta^\mu)&=&\frac{y^2}{ \cosh\eta^2}- y^1 \tanh \eta^1\tanh \eta^2\\
B^3(y^\mu,\eta^\mu)&=&\frac{y^3}{ \cosh\eta^3}- y^2\tanh\eta^2\tanh\eta^3- y^1\frac{\tanh \eta^1\tanh \eta^3}{\cosh\eta^2}\,.
\end{array}
\end{equation}
Note that the worldline $w_0$ corresponding to a particle at rest in the spacetime origin (see eq.~\eqref{eq:wl_0}) is described in the space of worldlines by $\eta^1=\eta^2=\eta^3=0$ and $y^1=y^2=y^3=0$.

We remark that there exists an alternative way to find the correspondence between the parametrization of a given worldline in the space of worldlines and in Minkowski spacetime. Let us consider the affine space defined by vectors of the form
\begin{equation}
\left(\begin{array}{c}
1\cr 
x^0 \cr 
x^1 \cr 
x^2 \cr 
x^3 
\end{array}\right) .
\end{equation}
The action of $G_\mathcal{M}$ on this affine space is linear and the identification with the previous description is straightforward. Namely, the privileged worldline $w_0$ can be defined as
\begin{equation}
w_0\equiv \left(\begin{array}{c}
1\cr 
x^0 \cr 
0 \cr 
0 \cr 
0 
\end{array}\right) =\left(\begin{array}{c}
1\cr 
\tau \cr 
0 \cr 
0 \cr 
0 
\end{array}\right),
\end{equation}
and any other worldline can be obtained by acting on it with the Poincar\'e group element given by 
\begin{equation}
g_w = \exp\{\xi^1 \rho(K_1)\} \exp\{\xi^2 \rho(K_2)\} \exp\{\xi^3 \rho(K_3)\} \exp\{\beta^1 \rho(P_1)\} \exp\{\beta^2 \rho(P_2)\} \exp\{\beta^3 \rho(P_3)\}\,.
\end{equation} 
In this way we get
\begin{equation}
w=\label{eq:wl_group}
g_w w_0 = g_w \left(\begin{array}{c}
1\cr 
\tau \cr 
0 \cr 
0 \cr 
0 
\end{array}\right)=\left(\begin{array}{c}
1\cr 
f^0(\beta^\mu,\xi^\mu) \cr 
f^1(\beta^\mu,\xi^\mu) \cr 
f^2(\beta^\mu,\xi^\mu) \cr 
f^3(\beta^\mu,\xi^\mu) 
\end{array}\right)\,,
\end{equation}
where the functions $f^\alpha$ are just the ones appearing in \eqref{eq:MWrelation} and  \eqref{eq:f}.
Now, by comparing \eqref{eq:wl_group} with \eqref{eq:wl_linearv} the relations \eqref{eq:vi}-\eqref{eq:Bi} are  found again.

\subsection{$\kappa$-Poisson structure and quantization}

Once the description of Minkowski spacetime ${\cal M}$ and of its associated space of time-like worldlines ${\cal W}$ has been given in the previous geometric setting, the $\kappa$-deformation can be introduced at the level of the Poincar\'e symmetries. As it was shown in~\cite{BGH2019worldlinesplb}, this is done by introducing an additional Poisson-Lie structure on the Poincar\'e group manifold, given by the classical $r$-matrix that underlies the $\kappa$-deformation:
\begin{equation}
\label{eq:r_kappapoincare}
r=\frac{1}{\kappa} (K_1 \wedge P_1 + K_2 \wedge P_2 + K_3 \wedge P_3) .
\end{equation}
The Poisson structure is then provided by the Sklyanin bracket associated to~\eqref{eq:r_kappapoincare}, which can be computed on the two parametrizations~\eqref{eq:Gm} and~\eqref{eq:Gw} of the Poincar\'e group by realizing the Lie algebra generators of the Poincar\'e algebra in terms of the corresponding left and right-invariant vector fields. Then, the two canonical projections of the Sklyanin Poisson structure on each of the homogeneous spaces ${\cal M}$ and ${\cal W}$ provide two Poisson structures which are covariant under the action of the Poisson-Lie Poincar\'e group in the appropriate parametrization. The quantization of these  Poisson structures  gives rise to the corresponding quantum spaces of points and worldlines, which are -by construction- covariant under the (co)action of the quantum $\kappa$-Poincar\'e group, whose algebra relations among its (now noncommutative) parameters are obtained as the quantization of the Sklyanin bracket.

More explicitly, the Poisson structure induced by~\eqref{eq:r_kappapoincare} on the classical Minkowski spacetime ${\cal M}$ is given in terms of the spacetime coordinates as
\begin{equation}
\label{eq:kspacetimePoisson}
\{x^0,x^a\}=- \frac{1}{\kappa} \, x^a, \qquad  \{x^a,x^b\}= 0\, ,
\end{equation}
and therefore we can say that the $\kappa$-deformation provides a (Poisson) noncommutative structure on ${\cal M}$. Moreover, the quantization of the Poisson algebra~\eqref{eq:kspacetimePoisson} gives rise to the well-known $\kappa$-Minkowski noncommutative spacetime
\begin{equation}
\label{eq:conmkappaMinkowski}
[\hat x^0,\hat x^a]=- \frac{1}{\kappa} \, \hat x^a, \qquad \qquad [\hat x^a,\hat x^b]= 0\, ,
\end{equation} 
where $\hat x^0$ and $\hat x^a$ are the operators that are assumed to encode  nontrivial localization properties, governed by the Planck mass parameter $\kappa$. We stress that in the limit $\kappa\to\infty$ both the Poisson structure~\eqref{eq:kspacetimePoisson} and the commutators~\eqref{eq:conmkappaMinkowski} vanish, and therefore Planck-scale kinematical effects disappear.

A similar procedure can be  performed on the space of worldlines ${\cal W}$. As it was shown in~\cite{BGH2019worldlinesplb}, the Poisson structure induced by~\eqref{eq:r_kappapoincare} onto this 6-dimensional space parametrized by the worldline coordinates $(\eta^i,y^i)$ is explicitly given by 
\begin{equation}
\label{eq:kworldlinesPoisson}
\begin{split}
&\{y^1,y^2\}= \frac 1{\kappa} \left(y^2 \sinh \eta^1 - \frac{y^1 \tanh \eta^2}{\cosh \eta^3} \right) ,\\
&\{y^1,y^3\}=  \frac 1{\kappa}    \left(y^3 \sinh \eta^1 - y^1 \tanh \eta^3 \right) , \\
&\{y^2,y^3\}=   \frac 1{\kappa}    \left(y^3 \cosh \eta^1 \sinh \eta^2 - y^2 \tanh \eta^3\right) ,\\
&\{y^1,\eta^1\}=  \frac 1{\kappa}  \frac{ \left(  \cosh \eta^1  \cosh \eta^2 \cosh \eta^3-1 \right)}{\cosh \eta^2 \cosh \eta^3},\\
&\{y^2,\eta^2\}=  \frac 1{\kappa}   \frac{ \left( \cosh \eta^1 \cosh \eta^2\cosh \eta^3 -1\right)}{\cosh \eta^3}  ,\\
&\{y^3,\eta^3\}=   \frac 1{\kappa}  \left( \cosh \eta^1 \cosh \eta^2 \cosh \eta^3-1 \right) ,\\
& \{y^a,\eta^b\}= 0,\quad a\ne b,   \qquad  \{\eta^a,\eta^b\}=  0 \, .
\end{split}
\end{equation}

Despite their apparent complexity, the Poisson brackets~\eqref{eq:kworldlinesPoisson} can be easily quantized and transformed into commutators defined in terms of the quantum worldline coordinates given by the operators $(\hat\eta^i,\hat y^i)$, since no ordering ambiguities emerge in this process. This leads to the noncommutative space of time-like worldlines, which is defined  as the nonlinear algebra: 
\begin{equation}
\label{eq:kworldlinesCommutators}
\begin{split}
&[\hat y^1, \hat y^2]= \frac 1{\kappa} \left( \hat  y^2 \sinh \hat \eta^1 - \frac{\hat y^1 \tanh \hat \eta^2}{\cosh \hat \eta^3} \right) ,\\
&[\hat y^1,\hat y^3]=  \frac 1{\kappa}    \left(\hat y^3 \sinh \hat \eta^1 - \hat y^1 \tanh \hat \eta^3 \right) , \\
&[\hat y^2,\hat y^3]=   \frac 1{\kappa}    \left(\hat y^3 \cosh \hat \eta^1 \sinh \hat \eta^2 - \hat y^2 \tanh \hat \eta^3\right) ,\\
&[\hat y^1,\hat \eta^1]=  \frac 1{\kappa}  \frac{ \left( \cosh \hat \eta^1 \cosh \hat \eta^2 \cosh \hat \eta^3-\mathbb{I} \right)}{\cosh \hat \eta^2 \cosh \hat \eta^3},\\
&[\hat y^2,\hat \eta^2]=  \frac 1{\kappa}   \frac{ \left( \cosh \hat \eta^1 \cosh \hat \eta^2\cosh \hat \eta^3 - \mathbb{I}\right)}{\cosh \hat \eta^3}  ,\\
&[\hat y^3,\hat \eta^3]=   \frac 1{\kappa}  \left( \cosh \hat \eta^1 \cosh \hat \eta^2 \cosh \hat \eta^3-\mathbb{I} \right) ,\\
&[\hat y^a,\hat \eta^b]= 0,\quad a\ne b,   \qquad  [\hat \eta^a,\hat \eta^b]=  0 \, .
\end{split}
\end{equation}
In~\cite{BGH2019worldlinesplb} it was shown that the Poisson algebra \eqref{eq:kworldlinesPoisson} strongly simplifies when written in terms of new coordinates $(q^{i},\,p^{i})$ defined as follows:
\begin{eqnarray}
q^1&=&\frac{\cosh \eta^2 \cosh \eta^3}{\cosh\eta^1 \cosh \eta^2 \cosh \eta^3-1}y^1\nonumber\\
q^2&=&\frac{ \cosh \eta^3}{\cosh\eta^1 \cosh \eta^2 \cosh \eta^3-1}y^2\nonumber\\
q^3&=&\frac{1}{\cosh\eta^1 \cosh \eta^2 \cosh \eta^3-1}y^3\nonumber\\
p^i&=&\eta^i\,. \label{eq:qpdefinition}
\end{eqnarray}
This interesting result stems from the fact that the even-dimensional Poisson structure~\eqref{eq:kworldlinesPoisson} is non-degenerate in the dense submanifold $(\eta_1,\eta_2,\eta_3)\neq (0,0,0)$. Therefore Darboux theorem guarantees that locally there exists a transformation, which in this case is given by~\eqref{eq:qpdefinition}, giving rise to the symplectic Poisson brackets of a standard  phase space, namely:
\begin{equation}
\label{eq:kworldlinesPoissonqp}
\{q^a,q^b\}= \{p^a,p^b\}= 0, \qquad   \{q^a,p^b\}= \frac 1 {\kappa} \, \delta_{ab}  .
\end{equation}
 Therefore, the quantization of the Poisson structure on ${\cal W}$ becomes particularly simple in terms of the quantum coordinates $(\hat q^{i},\hat p^{i})$, and leads to a (linear) Heisenberg-Weyl algebra:
\begin{equation}
\label{eq:kworldlinesCommutatorsqp}
[\hat q^a,\hat q^b]= [\hat p^a,\hat p^b]= 0, \qquad   [\hat q^a,\hat p^b]= \frac i {\kappa} \, \delta_{ab} .
\end{equation}
As we will show in the following, noncommutative worldlines described through~\eqref{eq:kworldlinesCommutatorsqp} can be used to study the fuzzy effects induced by the $\kappa$-deformation.


\section{Quantum $\kappa$-worldlines}
\label{sec:quantumworldlines}

 Given that the operators $(\hat q^{i},\, \hat p^{i})$ satisfy a noncommutative Heisenberg-Weyl  algebra, with $\kappa^{-1}$ playing the role that is usually played by the Planck's constant $\hbar$, we can take advantage of the well-known results of quantum mechanics and define a realization of the algebras \eqref{eq:kworldlinesCommutators} and \eqref{eq:kworldlinesCommutatorsqp} on a suitable Hilbert space. This allows us to define quantum states in such space and study the properties of observables associated to the worldlines coordinates, namely $(\hat y^{i}, \hat \eta^{i})$ and $(\hat v^{i},\hat B^{i})$. 

A word of caution: we stress that $\hbar$ does not appear anywhere in our model, and its role is played instead by $\kappa^{-1}$, which is now the parameter that governs the novel noncommutative structures induced by the existence of a deformed Poincar\'e symmetry. Having a noncommutative algebra, it is tempting to interpret its commutators as expressing incompatibility between operators associated to measurement operations, and to borrow the whole interpretative framework of quantum mechanics (uncertainty relations, probabilistic interpretation, Born rule, etc.). We must, however, warn the reader that our model does not deal with quantum effects understood as the dynamics of physical systems defined in the usual phase space of quantum mechanics, but rather it aims at describing the non-classical geometry of a space of worldlines. Using the interpretative framework of Heisenberg's and Schr\"odinger's formalisms is not, therefore, a necessity. We will, in light of this, be careful distinguishing between true quantum mechanics models ruled by the parameter $\hbar$, from \emph{noncommutative} models, which are the ones arising from the `quantization' of a given Poisson manifold such as~\eqref{eq:kworldlinesPoisson} in terms of the parameter $\kappa^{-1}$. With `quantization', in this latter case, we just mean that we make a correspondence between real coordinates on the chosen Poisson manifold and self-adjoint operators on a suitable Hilbert space.
This said, it is clear that the interpretative building of quantum mechanics offers a powerful tool to translate mathematical statements into physical predictions, and nothing prevents us from importing it into our models to get a physical interpretation, as long as the reader has been warned about the caveats. 

For simplicity, in this section we will work in (1+1) dimensions. The results are easily generalizable to a higher number of dimensions, and we will do so in the following section.

\subsection{Squeezed states}

Let us introduce a standard representation for the algebra~(\ref{eq:kworldlinesCommutatorsqp}) in the one-dimensional case. We start by defining quantum states that minimize the uncertainty. Recall that in (1+1) dimensions
\begin{equation}
[\hat q,\hat p]=\frac{i}{\kappa}\,,\label{eq:qpComm}
\end{equation}
so we can use the tools of standard quantum mechanics, with the replacement $\hbar \leftrightarrow \kappa^{-1}$: the operators $(\hat q,\hat p)$ are conjugate ones acting on states in a similar way as the usual quantum-mechanical coordinate and momentum operators, namely
\begin{equation}
\begin{split}
&\hat q\,  \psi(p) = \frac{i}{\kappa} \frac{\partial \psi(p)}{\partial p} \,,\\
&\hat p \, \psi(p) = p\, \psi(p).
\end{split}
\end{equation}
The Hilbert space is $\mathcal{H}=  L^2(\mathbbm{R})$ and the inner product between two states $\psi$ and $\varphi$ is given by
\begin{equation}
\langle \varphi | \psi \rangle = \int_{-\infty}^{\infty} dp \; \overline{\varphi}(p) \psi(p) \, .
\label{eq:innerproduct}
\end{equation}
We can consider the well-known squeezed states given  by
\begin{equation}
\psi (p)
=\frac{1}{\sigma^{1/2} \pi^{1/4}} e^{-\frac{(p-p_0)^2}{2 \sigma^2}} e^{-i \kappa q_0 p},\label{eq:squeezed_state}
\end{equation}
which reduce to non-squeezed coherent states when $\sigma =1$. We can verify explicitly that  the state \eqref{eq:squeezed_state} minimizes the uncertainty in  $(q,p)$. In order to do so, we compute the variance of the operators $(\hat q,\hat p)$. For the operator $\hat p$ we have:
\begin{equation}
\langle \hat p \rangle = \frac{1}{\sigma \pi^{1/2}} \int_{-\infty}^{\infty} dp \; p \; e^{-\frac{(p-p_0)^2}{\sigma^2}} = p_0\,,
\end{equation}
\begin{equation}
\langle \hat p^2 \rangle = \frac{1}{\sigma \pi^{1/2}} \int_{-\infty}^{\infty} dp \; p^2 \; e^{-\frac{(p-p_0)^2}{\sigma^2}} = p_0^2 + \frac{\sigma^2}{2}\,,
\end{equation}
thus the variance of $\hat p$ is $\delta \hat p \equiv \sqrt{\langle \hat p^2 \rangle - \langle \hat p \rangle^2} = \frac{\sigma}{\sqrt 2}$. For the operator $\hat q$ one finds:
\begin{equation}
\begin{split}
&\langle \hat q \rangle = \frac{1}{\sigma \pi^{1/2}} \int_{-\infty}^{\infty} dp \; (e^{-\frac{(p-p_0)^2}{2 \sigma^2}} e^{-i \kappa q_0 p}) ( \frac{i}{\kappa} \frac{\partial}{\partial p}) (e^{-\frac{(p-p_0)^2}{2 \sigma^2}} e^{i \kappa q_0 p}) \\
&=\frac{1}{\sigma \pi^{1/2}} \int_{-\infty}^{\infty} dp \; e^{-\frac{(p-p_0)^2}{\sigma^2}} q_0 = q_0\,,
\end{split}
\end{equation}
\begin{equation}
\begin{split}
&\langle \hat q^2 \rangle = \frac{1}{\sigma \pi^{1/2}} \int_{-\infty}^{\infty} dp \; (e^{-\frac{(p-p_0)^2}{2 \sigma^2}} e^{-i \kappa q_0 p}) (-\frac{1}{\kappa^2} \frac{\partial^2}{\partial p^2}) (e^{-\frac{(p-p_0)^2}{2 \sigma^2}} e^{i \kappa q_0 p}) = q_0^2 + \frac{1}{2 \kappa^2 \sigma^2}\,,
\end{split}
\end{equation}
thus the variance of $\hat q$ is $\delta \hat q \equiv \sqrt{\langle \hat q^2 \rangle - \langle \hat q \rangle^2} = \frac{1}{\sqrt 2 \sigma \kappa}$. Then the product of the uncertainties is:
\begin{equation}
\label{eq:deltapq}
\delta \hat p \, \delta \hat q= \frac{\sigma}{\sqrt 2} \frac{1}{\sqrt 2 \sigma \kappa}= \frac 1 2 \kappa^{-1}\,,
\end{equation}
which corresponds indeed to the minimum uncertainty associated to the commutation relation \eqref{eq:qpComm}. 

While the operators $(\hat q,\hat p)$ have simple commutation relations and are thus  convenient to use in order to define states, their physical relevance is limited. As we discussed in the previous sections, the physically meaningful operators are either $(\hat y^{i}, \hat \eta^{i})$, which are the noncommutative coordinates in the space of worldlines, or $(\hat v^{i},\hat B^{i})$, which identify worldlines in spacetime. In the following we will see that the states~\eqref{eq:squeezed_state} do not minimize the uncertainty product for the operators  $(\hat y, \hat \eta)$ in the (1+1)-dimensional case. To do so, we introduce these operators by inverting the (1+1)-dimensional equivalent of equations \eqref{eq:qpdefinition}  and symmetrizing:
\begin{equation}\label{eq:RepresentationEtaY}
\begin{aligned}
\hat \eta &= \hat p \,, \\
\hat y &=  \, : (\cosh \hat p - 1) \hat q : \, = (\cosh \hat p - 1) \hat q + \frac{i}{2 \kappa} \sinh \hat p = \frac{i}{\kappa} (\cosh p -1) \frac{\partial}{\partial p} + \frac{i}{2 \kappa} \sinh p\,.
\end{aligned}
\end{equation}

Using the above definition, one can compute the variances on the squeezed state \eqref{eq:squeezed_state}. We start by computing the expectation values of the operators and their square, which read:
\begin{equation}
\langle \hat \eta \rangle = p_0,
\end{equation}
\begin{equation}
\langle \hat \eta^2 \rangle = p_0^2 + \frac{\sigma^2}{2},
\end{equation}
\begin{equation}
\langle \hat y \rangle = q_0 \big( e^{\frac{\sigma^2}{4}} \cosh p_0 - 1 \big),
\end{equation}
\begin{equation}
\begin{aligned}
\langle \hat y^{2} \rangle &= \frac{e^{-2 p_0}}{16 \kappa ^2 \sigma ^2}
 \bigg(-8 \left(e^{2 p_0}+1\right) e^{p_0+\frac{\sigma ^2}{4}} \left(2 \kappa ^2 q_0^2 \sigma ^2+1\right)+\\
 &\left(e^{4 p_0}+1\right) e^{\sigma ^2} \left(\sigma ^2 \left(4 \kappa ^2 q_0^2+1\right)+2\right)+2 e^{2 p_0} \left(\sigma ^2 \left(12 \kappa ^2 q_0^2-1\right)+6\right)\bigg).
\end{aligned}
\end{equation}
From these, we find that the variances are
\begin{equation}
\begin{aligned}
\delta \hat \eta&=   \frac{\sigma }{\sqrt{2}}\,, \\
\delta \hat y &= 
\bigg[
 \frac{e^{-2 p_0}}{16 \kappa^2 \sigma^2}\bigg(-8 \left(e^{2 p_0}+1\right) e^{p_0+\frac{\sigma ^2}{4}} +\\
&\left(e^{4 p_0}+1\right) e^{\sigma ^2} \left(\sigma ^2 \left(4 \kappa ^2 q_0^2+1\right)+2\right)+2 e^{2 p_0} \left(\sigma ^2 \left(12 \kappa ^2 q_0^2-1\right)+6\right)\bigg)- 
q_0^2 \left(1- e^{\frac{\sigma ^2}{4}} \cosh (p_0)\right)^2
\bigg]^{1/2}\, .
\end{aligned}
\end{equation}

Since $\hat y$ and $\hat \eta$ are self-adjoint operators by construction, it is guaranteed that this product is always greater than the Heisenberg bound derived from the quantum commutation relations~(\ref{eq:kworldlinesCommutators}), since in (1+1) dimensions the algebra (\ref{eq:kworldlinesCommutators}) reduces to:
\begin{equation}
[\hat y , \hat \eta ] = \frac i \kappa \left( \cosh \hat \eta - 1 \right) \,.
\end{equation}
This commutator implies uncertainty relations of the form
\begin{equation}
\delta \hat y \,\delta \hat \eta \geq \frac 1 {2\kappa}\left| \langle \cosh \hat \eta - 1 \rangle\right| \,.\label{eq:Hyeta}
\end{equation}
On the states \eqref{eq:squeezed_state} the right hand side takes the value
\begin{equation}
\frac 1 {2\kappa} \left| \langle \cosh \hat \eta - 1 \rangle \right| = \frac 1 {2\kappa} \left( e^{\sigma^{2}/4} \cosh p_{0} - 1\right)\,,
\end{equation}
and  it is easy to explicitly check the inequality \eqref{eq:Hyeta} by noting that on the states \eqref{eq:squeezed_state}
\begin{equation}
\begin{aligned}
(\delta \hat y \,\delta \hat \eta)^2 - \frac{1}{4}( \langle[ \hat \eta, \hat y ] \rangle )^2 = \\
\frac{1}{16 \kappa ^2} \bigg(-4 e^{\frac{\sigma ^2}{2}} &\cosh ^2(p_0) \left(2 \kappa ^2 q_0^2 \sigma ^2+1\right)+e^{\sigma ^2} \cosh (2 p_0) \left(4 \kappa ^2 q_0^2 \sigma ^2+\sigma ^2+2\right)+\sigma ^2 \left(4 \kappa ^2 q_0^2-1\right)+2\bigg) \, ,
\end{aligned}
\end{equation}
which  is always positive for all $\sigma > 0$. We thus conclude that the states \eqref{eq:squeezed_state}, while saturating the Heisenberg  bound for the operators $(\hat q, \hat p)$ (see \eqref{eq:deltapq}), do not do so for the operators $(\hat y,\hat \eta)$.

\subsection{Perfectly localized state in the origin of the space of worldlines}
\label{perfecly_localized_state}

While the squeezed states \eqref{eq:squeezed_state} do not saturate the Heisenberg uncertainty bound for the operators $(\hat y, \hat \eta)$, the uncertainty relations \eqref{eq:Hyeta} allow for perfectly localized states. In fact, computed on states such that  $\langle \cosh \hat \eta \rangle = 1$, the right-hand side of  \eqref{eq:Hyeta} is zero. In this subsection we show that these states can be defined as limits of normalized states, as was done in~\cite{LMMP2018localization,arXiv:1912.07098} for the states localized at the spatial origin in $\kappa$-Minkowski spacetime. In this way, we are able to define a state in the origin of the space of worldlines, corresponding to the worldline $w_{0}$, that can be used as a sharp reference to compute the impact parameter \eqref{eq:distw0explicit} of a quantum worldline $w$ with respect to it. It is important that the reference worldline $w_{0}$ can be defined sharply because of the issues with Lorentz invariance that would emerge otherwise, see section \ref{sec:impact}.

We begin by observing that the variable $\hat \rho  = \kappa \coth (\hat \eta/2)$ is canonically conjugate to $\hat y$:\footnote{Notice that $\hat \rho$ is different from the variable $\hat q$ showed above, since $\hat\rho$ is a function of $\hat \eta$ alone, while $\hat q$  is a function of $\hat \eta$ and  $\hat y$.}
\begin{equation}
[ \hat y ,\hat \rho ] = i \,,
\end{equation}
and so, in the basis of eigenstates of $\hat \rho$, a Gaussian function would saturate the uncertainty product $\delta \hat y\, \delta \hat \rho$ and be a promising start for a localized state. The problem is that, in terms of the eigenstates of $\hat \eta$, such a wavefunction takes the form:
\begin{equation}
\psi = N e^{- \frac{\kappa^2\left( \coth \frac \eta 2 - \coth \frac{\eta_0}{2}\right)^2}{2 \sigma^2}} \,, 
\end{equation}
and, as $\eta \to \infty$, this wavefunction tends to a constant (because $\coth \eta/2 \xrightarrow[\eta \to \infty]{} 1$). Such a wavefunction is non-normalizable, because the inner product~(\ref{eq:innerproduct}) makes its norm divergent.

However, we can cut off this wavefunction at large $\eta$'s, because we are interested in localizing around small $\eta$'s. Moreover, we are just interested in the behaviour of the wavefunction with $\eta_0 =0$ around $\eta =0$, where $\coth ( \eta /2) \sim \frac 2 {\eta}$. We can therefore take a wavefunction of the form
\begin{equation}
\psi_{a,b} = N e^{- \frac{1+b \eta^4}{2 a \eta^2}} \,,
\end{equation}
which is normalizable, with
\begin{equation}
N = \left(\int_{-\infty}^{+\infty} \left| \psi_{a,b} \right|^2 d\eta \right)^{-1/2} =  \left(\frac{\pi a }{b} \right)^{-1/4} e^{\frac{ \sqrt b}{a}}\,.
\end{equation}
The expectation value and squared variance of $\hat \eta$ on this state are:
\begin{equation}
\langle \hat \eta \rangle = 0 \,,
\qquad 
\langle \hat \eta^2 \rangle = \frac{a}{2b} + \frac{1}{\sqrt{b}} \,,
\end{equation}
and they both go to zero as $b \to + \infty$ (as long as $a$ grows more slowly than $b$).

Now, we want to calculate the expectation value and variance of $\hat y$ onto the normalized state $\psi_{a,b}$, where $\hat y$ is represented as the following self-adjoint differential operator:
\begin{equation}
\hat y = \frac{i}{\kappa} \left( \cosh \eta - 1 \right) \frac{\partial}{\partial \eta} + \frac{i}{2\kappa} \sinh \eta \,.
\end{equation}
The expectation value $\langle \hat y \rangle $ is zero by antisymmetry. Setting $a = \sqrt b$, we find the following expression for the squared variance:
\begin{equation}
\begin{aligned}
\langle \hat y^2 \rangle =&
\int_{-\infty}^{+\infty} \frac{e^{2-\frac{b \eta ^4+1}{\sqrt{b} \eta ^2}} \sinh ^2\left(\frac{\eta }{2}\right)}{2 \sqrt{\pi } b^{3/4} \eta ^6} \big{[}
4 b^2 \eta ^8 -4 b^{3/2} \eta ^6
-b \eta ^6-8 b \eta ^4-12 \sqrt{b} \eta ^2+4
\\&
+\left(4 b^{3/2} \eta ^6-4 b^2 \eta ^8-3 b \eta ^6+8 b \eta ^4+12 \sqrt{b} \eta ^2-4\right) \cosh \eta  +8 \sqrt{b} \eta ^3 \left(b \eta ^4-1\right) \sinh \eta
\big{]} d\eta \,.
\end{aligned}
\end{equation}
This integral is finite for any positive $b$ and goes to zero as $b \to + \infty$ like $b^{-1/2}$. In fact, changing variable to $\eta = \frac{x}{b^{1/4}}$, the integrand can be approximated as
\begin{equation}
\langle \hat y^2 \rangle = b^{-1/2} 
\int_{-\infty}^{+\infty} \frac{e^{2-\frac{x^4+1}{x^2}} \left(x^{12}-5 x^{10}+x^6+x^4 + \mathcal O(b^{-1/2}) \right)}{2 \left(2 \sqrt{\pi } x^6\right)} dx = \frac{37}{32} b^{-1/2} + \mathcal O(b^{-1})  \,.
\end{equation}
Therefore, we have found a one-parameter family of normalized wavefunctions, which tend to a state perfectly localized in $\eta = y = 0$. A similar result can be proven for a state perfectly localized in 
$\eta =  0$, $y = y_0 \neq 0$.

\subsection{From quantum to semiclassical approach}

In the following section we will study the properties of `semiclassical worldlines', understood as probability distributions in the space of worldlines. Not all such distributions can  legitimately be called `semiclassical': many violate, for example, the quantum uncertainty bound.
There is however a class of probability distributions that inherit the properties of quantum states such as those studied in the current section. In order to construct them we follow the methodology introduced by Wigner~\cite{Wigner1932} to relate the quantum wavefunction to a statistical distribution in phase space: the \emph{quasiprobability distribution}.

In (1+1) dimensions the Wigner quasiprobability distribution associated to the quantum state $\psi (p)$ is defined as:
\begin{equation}
W(p,q) \equiv \frac{\kappa}{\pi} \int_{-\infty}^{\infty} dy \; \overline{\psi} (p+y) \psi(p-y) e^{-2 i \kappa y q} .
\end{equation}
Thus, in the case of the squeezed state~\eqref{eq:squeezed_state} the quasiprobability is given by
\begin{equation}
W(p,q) = \frac{\kappa}{\pi} e^{-\frac{(p-p_0)^2}{\sigma^2}} e^{- \kappa^2 \sigma^2 (q-q_0)^2}.
\end{equation}
This is a 2D normal distribution: 
\begin{equation}
F(q,p) = \frac{1}{2 \pi \sigma_q \sigma_p} e^{-\frac{1}{2}\left(\frac{(q-q_0)^2}{\sigma_q^2} + \frac{(p-p_0)^2}{\sigma_p^2}\right)}\label{eq:2DNormal}
\end{equation}
with $\sigma_q = \frac{1}{\sqrt{2} \sigma \kappa}$ and $\sigma_p = \frac{\sigma}{\sqrt{2}}$ satisfying by construction the Heisenberg uncertainty bound.

Since we are ultimately interested in studying intersections of worldlines, after a first exploratory investigation in (1+1) dimensions, we will then focus on (2+1) spacetime dimensions, because in (1+1) dimensions all non-parallel worldlines intersect. In fact, the function $\beta$ from Eq.~\eqref{eq:distw0explicit} is nontrivial only in (2+1) dimensions or higher.
For this reason, we will need the  (2+1)-dimensional Wigner quasiprobability distribution, which  is defined by 
\begin{equation}
W(p,q) \equiv \left( \frac{\kappa}{\pi} \right)^2 \int_{-\infty}^{\infty} \int_{-\infty}^{\infty} dy^1 \; dy^2 \; \overline{\psi} (p^1+y^1,p^2+y^2) \psi(p^1-y^1,p^2-y^2) e^{-2 i \kappa y^1 q^1} e^{-2 i \kappa y^2 q^2} \,.
\end{equation}
In the case of an isotropic Gaussian wavepacket:
\begin{equation}
\psi(p^1,p^2) = \frac{1}{\sigma \pi^{1/2}} e^{-\frac{(p^1-p^1_0)^2}{2 \sigma^2}} e^{-\frac{(p^2-p^2_0)^2}{2 \sigma^2}} e^{-i \kappa q^1_0 p^1} e^{-i \kappa q^2_0 p^2} \,,
\end{equation}
the quasiprobability distribution reduces to:
\begin{equation}\label{eq:WignerGaussian2D}
W(p,q) = \frac{\kappa^2}{\pi^2} e^{-\frac{(p^1-p^1_0)^2 +(p^2-p^2_0)^2}{\sigma^2}} e^{- \kappa^2 \sigma^2 \left[ (q^1-q^1_0)^2 + (q^2-q^2_0)^2 \right]}.
\end{equation}

This is a 4D normal distribution: 
\begin{equation}
F(q^1,q^2,p^1,p^2) = \frac{1}{4 \pi^2 \sigma_{q^1} \sigma_{q^2} \sigma_{p^1} \sigma_{p^2}} e^{-\frac{1}{2}\left(\sum_{i=1}^2 \frac{(q^i-q^i_0)^2}{\sigma_{q^i}^2} + \sum_{i=1}^2 \frac{(p^i-p^i_0)^2}{\sigma_{p^i}^2}\right)}\label{eq:4DNormal}
\end{equation}
with $\sigma_{q^i} = \frac{1}{\sqrt{2} \sigma \kappa}$ and $\sigma_{p^i} = \frac{\sigma}{\sqrt{2}}$ satisfying by construction the Heisenberg uncertainty bound.

\section{Fuzzy $\kappa$-worldlines}
\label{Sec:FuzzyWorldlines}

In this section we explore the properties of the space of worldlines that are implied by the non-trivial commutators \eqref{eq:kworldlinesCommutators} between the coordinate operators on this space $(\hat y^{i},\hat \eta^{i})$. As we mentioned, 
our aim is that of making a first step in the direction of reconstructing spacetime as the set of events defined by the intersection of different worldlines. In particular, we want to study the induced fuzziness of these events due to the quantum nature of the $\kappa$-worldlines. 

At the classical level, the intersection of two worldlines is defined via the observable $\beta$, defined in section \ref{sec:classicalworldlines}, eq.~\eqref{eq:distw0explicit}. This observable takes zero value  when two worldlines cross. As we explained in section \ref{sec:impact}, we can generalize this observable to the quantum setting if the worldline $w_{0}$ is that of a particle sharply at rest in the origin, corresponding to sharply vanishing expectation values of $(\hat y,\hat \eta)$.

In the previous section we have seen that these operators are in general characterized by an irreducible uncertainty $\delta \hat y\,\delta \hat \eta>0$, unless they are evaluated on a special state corresponding to the worldline of a particle at rest. In fact, as shown in subsection \ref{perfecly_localized_state}, this state can be perfectly localized, so that  it describes a worldline with sharp values of the parameters $(\hat y=y_{0},\hat \eta=0)$, and in particular one can choose $(\hat y=0,\hat \eta=0)$, corresponding to the worldline $w_{0}$. This allows us to define the quantum $\hat \beta$ operator.

In principle, in order to study the properties of $\hat\beta$ at the quantum level, we would be required to perform quantum tomography on the wavefunctions representing the state of our system (which is essentially the wavefunction of the worldline that is not at rest in the origin). This is however unpractical, in particular if we are interested in (semi-)analytical results, considering the complicated form of $\hat \beta$ in the representation~\eqref{eq:RepresentationEtaY}. We can, however, get heuristic results that accurately reproduce the statistical properties of quantum states if we work at the semiclassical (\emph{i.e.}~commutative-but-non-Poisson commutative) level, with classical probability distributions. This is legitimate as long as we use the probability distribution corresponding to the  Wigner quasiprobability of a legitimate quantum wavefunction.
 
The Gaussian distributions found in the previous section will then be the starting point of the analysis in this section.  We will be able to work out the induced distribution of the coordinates $(y^{i},\eta^i)$ and of the worldlines parameters $(v^{i}, B^{i})$ by using the relations \eqref{eq:qpdefinition} and \eqref{eq:vi}-\eqref{eq:Bi}, respectively. Finally, we will analyze the distribution of the impact parameter $\beta$ defined in section \ref{sec:impact}, getting insights about the properties of `fuzzy' events.
 
\subsection{Distribution of worldlines in (1+1) dimensions}

In order to illustrate our method of analysis, we start by studying the distribution corresponding to a squeezed state of a (1+1) dimensional worldline. 
If the distribution is centered  in $(p,q)=(0,0)$, then it takes the form already found in \eqref{eq:2DNormal}.
The relation between the  variables $(q,p)$ and the worldline coordinates $(y,\eta)$ is given in~\eqref{eq:qpdefinition}. In (1+1) dimensions this takes the form: 
\begin{equation}
(q,p)  \longrightarrow \left( \frac{y}{\cosh \eta -1},\eta \right), \label{eq:alpha}
\end{equation}
defined whenever $\eta \neq 0$. The probability density in the $(y, \eta)$ plane, induced by the  distribution \eqref{eq:2DNormal} in $(p,q)$, is given by
\begin{equation}
G(y,\eta)  = \frac{1}{2 \pi \sigma_q \sigma_p (\cosh \eta -1)} e^{-\frac{1}{2}\left(\frac{y^2}{\sigma_q^2 (\cosh \eta -1)^2} + \frac{\eta^2}{\sigma_p^2}\right)}\,,\label{eq:Gyeta}
\end{equation}
where we used the Jacobian of the transformation \eqref{eq:alpha} and $G$ is normalized in such a way that
\newline  $\int_{-\infty}^\infty \int_{-\infty}^\infty dy d\eta \,G(y,\eta) =1$. While in principle this distribution is not defined in $\eta=0$, it can be extended in a continuous way to the whole $(y,\eta)$ plane, including the $\eta=0$ line, where $G=0$.

\begin{figure}[htbp]
\begin{center}
\includegraphics[scale=0.6]{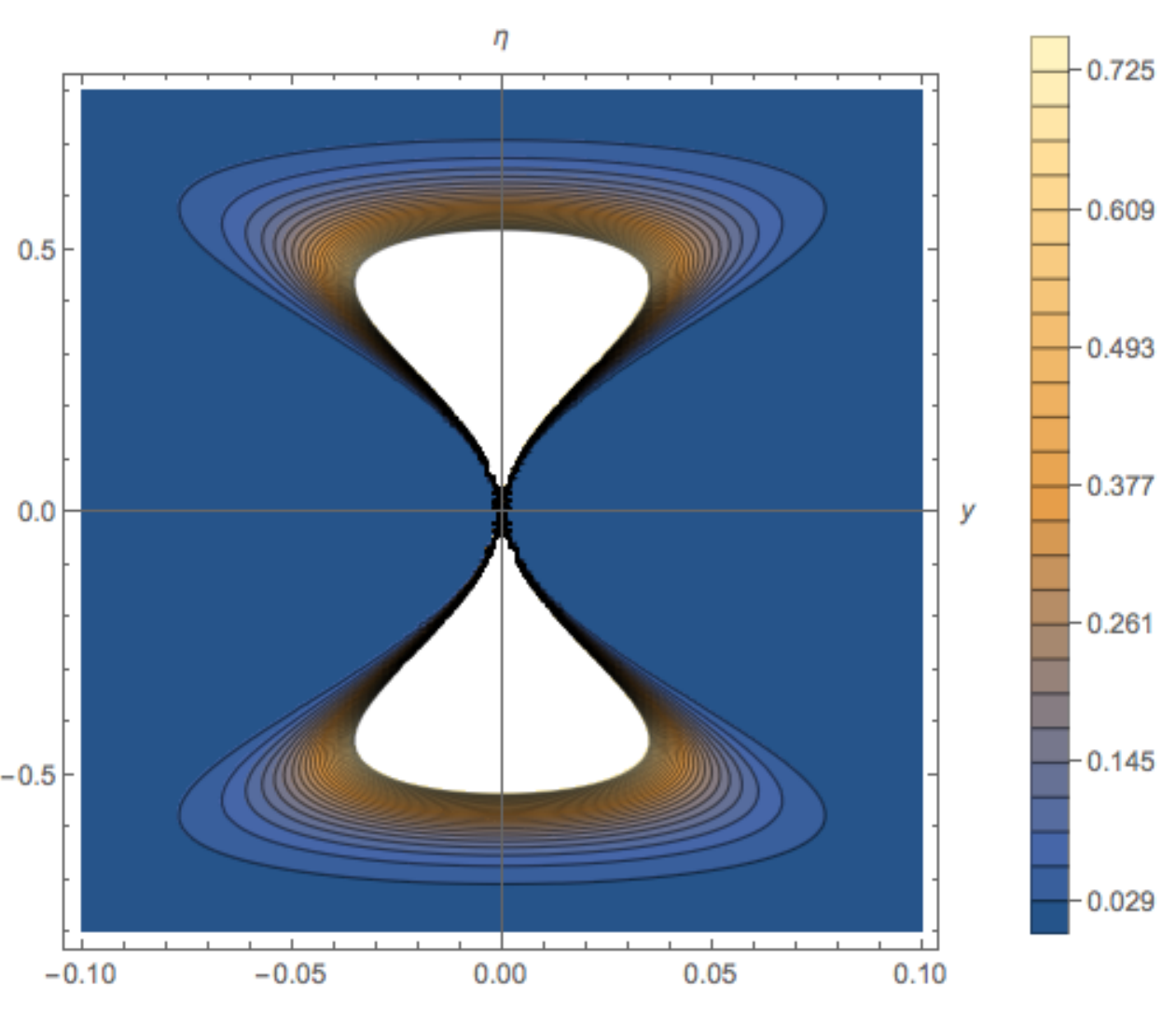}
\includegraphics[scale=0.6]{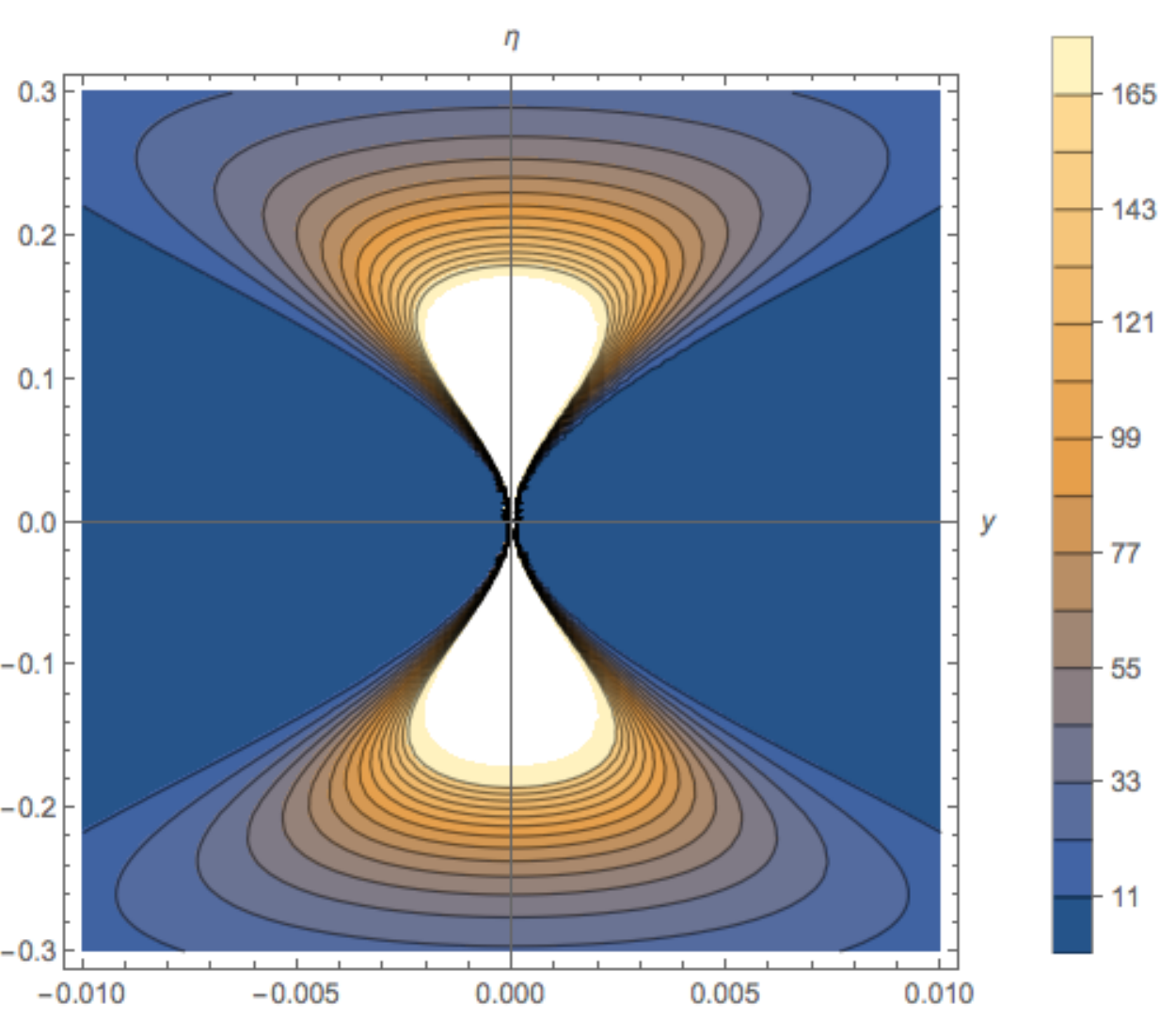}
\caption{Contour plots for the distribution $G(y,\eta)$, eq.\eqref{eq:Gyeta}, at two different scales, for  $\sigma_p = \sigma_q=0.2$. }
\label{fig:ftilde}
\end{center}
\end{figure}

\begin{figure}[htbp]
\begin{center}
\includegraphics[scale=0.65]{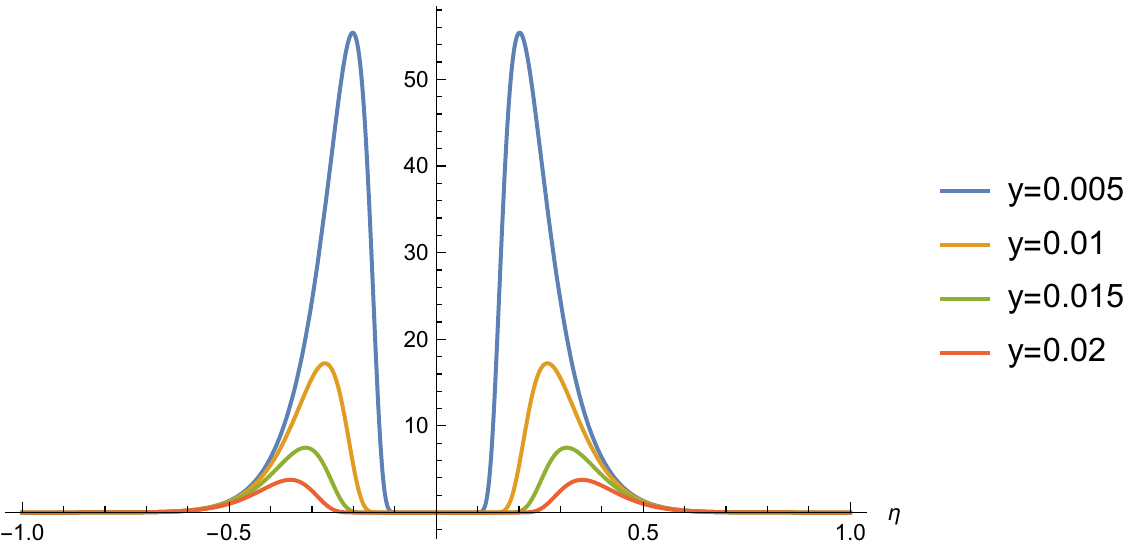}
\includegraphics[scale=0.65]{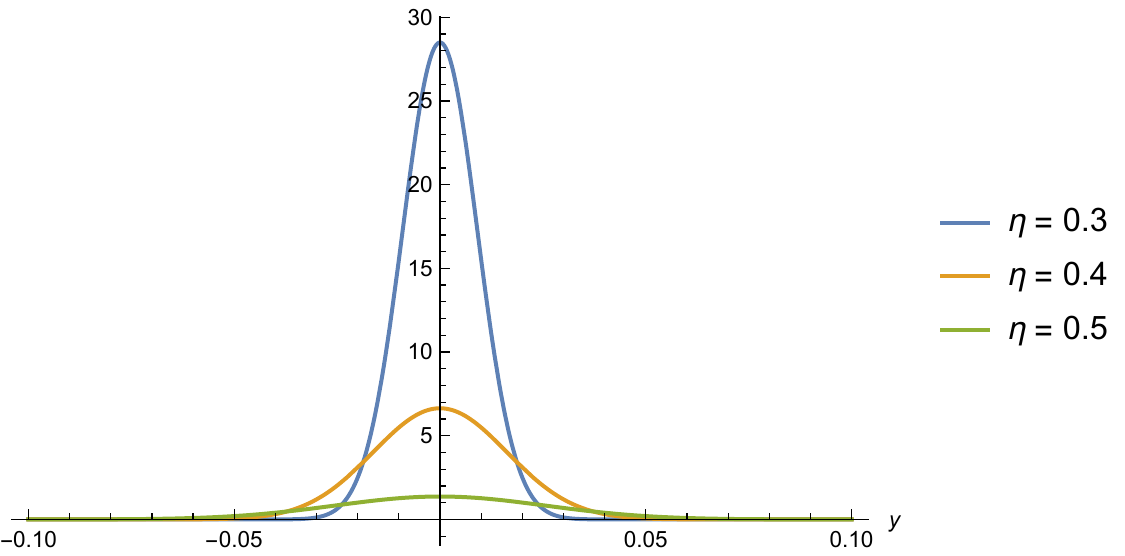}
\caption{Several sections $y=\text{constant}$ (left) and $\eta=\text{constant}$ (right) of the distribution $G(y,\eta)$, eq. \eqref{eq:Gyeta}, for $\sigma_p = \sigma_q=0.2$.  }
\label{fig:yConstantSection}\label{fig:etaConstantSection}
\end{center}
\end{figure}

The distribution $G$ is plotted in figures \ref{fig:ftilde} and \ref{fig:etaConstantSection}.
Some interesting information can be extracted by computing the marginal distributions for $y$ and $\eta$. 
The marginal distribution for $y$,
\begin{equation}
G(y) = \int_{-\infty}^\infty d \eta \, \frac{1}{2 \pi \sigma_q \sigma_p (\cosh \eta -1)} e^{-\frac{1}{2}\left(\frac{y^2}{\sigma_q^2 (\cosh \eta -1)^2} + \frac{\eta^2}{\sigma_p^2}\right)}  \, ,
\end{equation}
is a function sharply peaked in $y=0$, where it takes infinite value. On the other hand, the marginal distribution for $\eta$,
\begin{equation}
G(\eta) = \int_{-\infty}^\infty d y \, \frac{1}{2 \pi \sigma_q \sigma_p (\cosh \eta -1)} e^{-\frac{1}{2}\left(\frac{y^2}{\sigma_q^2 (\cosh \eta -1)^2} + \frac{\eta^2}{\sigma_p^2}\right)} =\left\{ \begin{array}{l}
0 \qquad\quad\text{if}\quad \eta=0\\
\frac{  e^{-\frac{\eta^{2}}{2\sigma_{p}^{2}}}  }{  \sqrt{2\pi}\sigma_{p}   } \quad\text{if}\quad \eta\neq 0,
\end{array}
\right.
\end{equation}
 is a Gaussian of variance $\sigma_{p}$, except in $\eta=0$, where it takes value zero.

\begin{figure}[h]
\begin{center}
\includegraphics[scale=0.4]{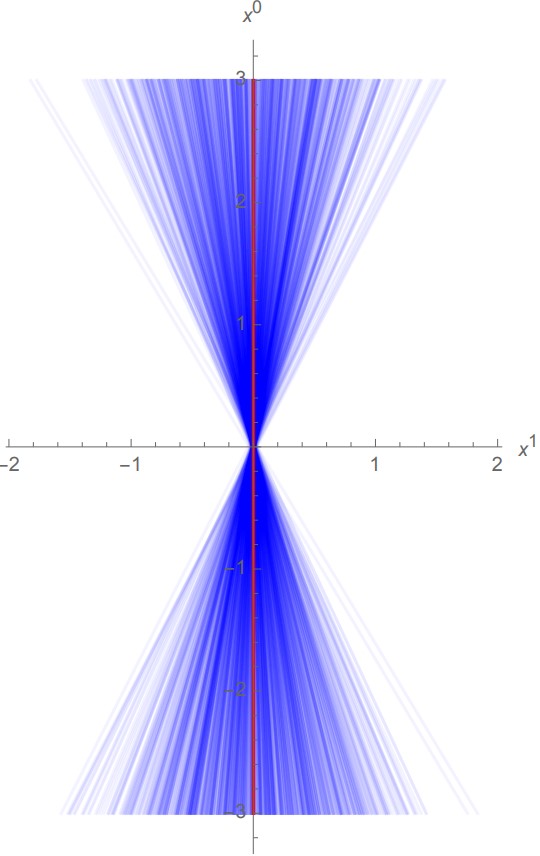}
\caption{Bundle of worldlines resulting from the extraction of 1000 points from the distribution $H(v,B)$ defined in~\eqref{eq:v_B_gaussian_zero}. The red worldline corresponds to the center of the distribution, with  $(p,q)=(0,0)$. Each worldline has an opacity of $0.05$.}\label{fig:wordlines_bundle_00}
\end{center}
\end{figure}

In order to grasp a more physically meaningful picture it is useful to look at the distribution in terms of the worldline parameters $(v,B)$, whose relation to $(y,\eta)$ is given in~\eqref{eq:vi}-\eqref{eq:Bi}. This allows us to plot the bundle of worldlines corresponding to the Gaussian distribution in the $(p,q)$ space we defined above, see figure \ref{fig:wordlines_bundle_00}.  

Using \eqref{eq:vi} and \eqref{eq:Bi} and specializing again to (1+1) dimensions, we have that the $(B,v)$ coordinates are related to the $y,\eta$ ones via:
\begin{equation}
(y,\eta) \longrightarrow \left( \frac{B}{\sqrt{1-v^2}},\frac{1}{2} \log\left(\frac{1+v}{1-v} \right) \right)\,.
\end{equation}
Therefore, the probability density function $H(B,v)$ takes the form 
\begin{equation}\label{eq:v_B_gaussian_zero}
H(B,v) =\frac{1}{2 \pi \sigma_p \sigma_q \left(v^2-1\right) \left(\sqrt{1-v^2}-B \right)} \left(\frac{1+v}{1-v}\right)^{\frac{1}{2 \sigma_p^2}} e^{-\frac{B^2}{2 \sigma_q^2 \left(B-\sqrt{1-v^2}\right)^2}}\,.
\end{equation}
This function is  very similar to the density function for $G (y, \eta)$. This can be easily understood since $G(y, \eta)$ is only different from zero when $\eta <<1$, and in this case $v \sim \eta$ and $B \sim y$. Therefore the properties of $H(B,v)$ and $G (y, \eta)$ are essentially the same.

As we mentioned, using the distribution of $(B,v)$ we can plot the bundle of worldlines corresponding to the distribution $H$. From~\eqref{eq:wl_linearv}, we have that in (1+1) dimensions a generic worldline reads
\begin{equation}
x^1=v \;x^0+B\,.
\end{equation}
Since $ v(p=0, q=0)=0$ and  $B(p=0,q=0)=0$, the worldline corresponding to the central point of the Gaussian distribution used above, $p=0,q=0$, is
 \begin{equation}
x^1=0,
\end{equation}
so this is the worldline $w_{0}$ corresponding to a stationary particle in the origin of Minkowski spacetime.
We can then randomly extract points from the distribution  $H(v,B)$ defined in eq. \eqref{eq:v_B_gaussian_zero} and associate a worldline to each of them. The bundle of worldlines resulting from the extraction of $1000$ random points is depicted in figure \ref{fig:wordlines_bundle_00}. 

Until now we studied the properties of a distribution corresponding to a squeezed state in $(p,q)$ centred in the origin. As we explained in the previous section, the origin in $(p,q)$ is in fact a special point, since it allows for a sharply localised distribution. So the distribution of worldlines centred around the worldline corresponding to a particle stationary in the origin does not need to be fuzzy. For this reason it makes sense to look at a distribution that is centred around a different worldline, since in that case the squeezed distribution is indeed the best localisation we can achieve.
 
To this aim, we repeat a similar procedure as above to study the bundle of worldlines generated by a Gaussian distribution in $(p,q)$ centered at a different point, for example $(\bar p,\bar q)=(1,1)$. The worldline corresponding to the central point of this distribution is 
 \begin{equation}
x^1=\tanh\bar\eta\; x^0+\frac{\bar y}{\cosh{\eta}},
\end{equation}
with
\begin{eqnarray}
\bar\eta &=& \eta(p=1)=1\,,\nonumber\\
\bar y&=& y(p=1,q=1)=\cosh{1}-1\, ,
\end{eqnarray}
or 
\begin{equation}
\begin{split}
\bar v&= v(p=1,q=1)=\tanh{1}\,, \\
\bar B&= B(p=1,q=1)=1-\frac{1}{\cosh 1}\, .
\end{split}
\end{equation}

\begin{figure}[h]
\begin{center}
\includegraphics[scale=0.35]{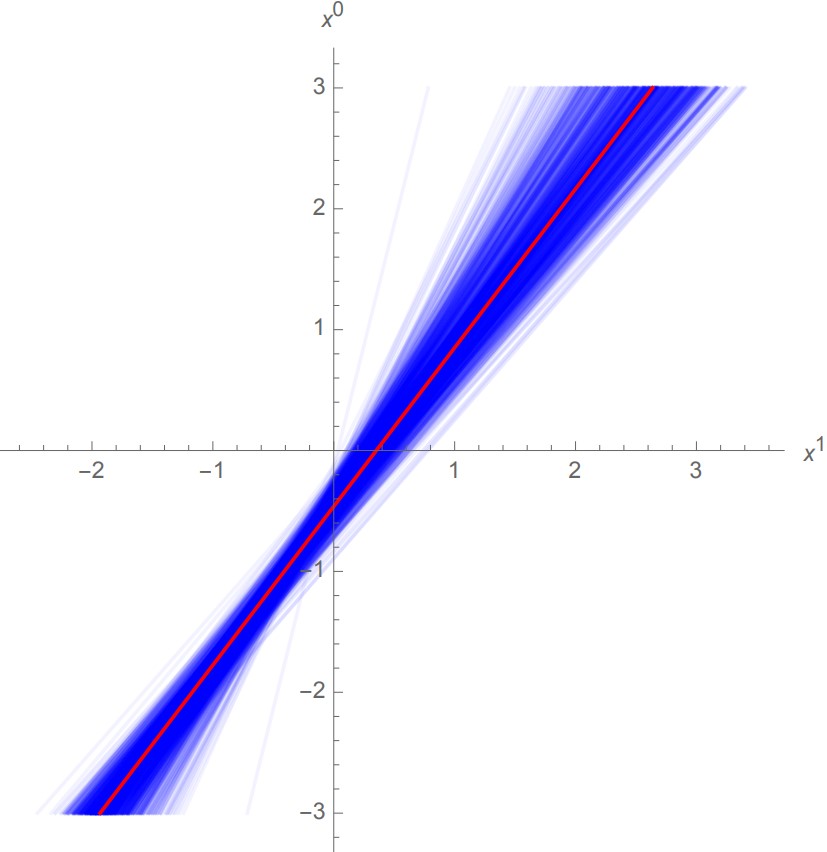}
\caption{Bundle of worldlines resulting from the random extraction of 1000 points from a Gaussian distribution in $(p,q)$ centered in $(p,q)=(1,1)$. The red worldline corresponds to the center of the distribution, with $(p,q)=(1,1)$. Each worldline has an opacity of $0.05$.}
\label{fig:wordlines_bundle_11}
\end{center}
\end{figure}

\begin{figure}[!h]
\begin{center}
\includegraphics[scale=0.55]{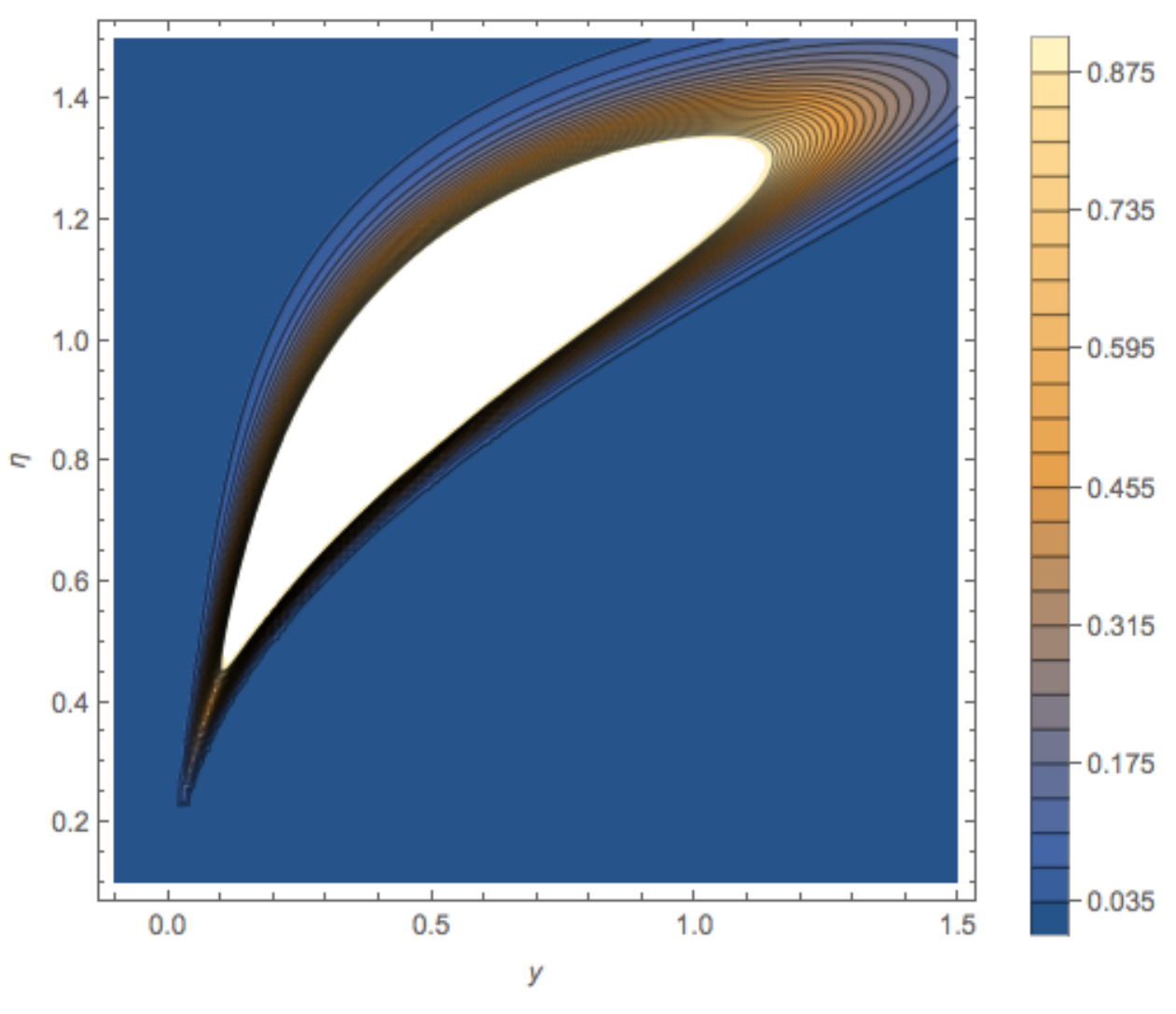}
\includegraphics[scale=0.55]{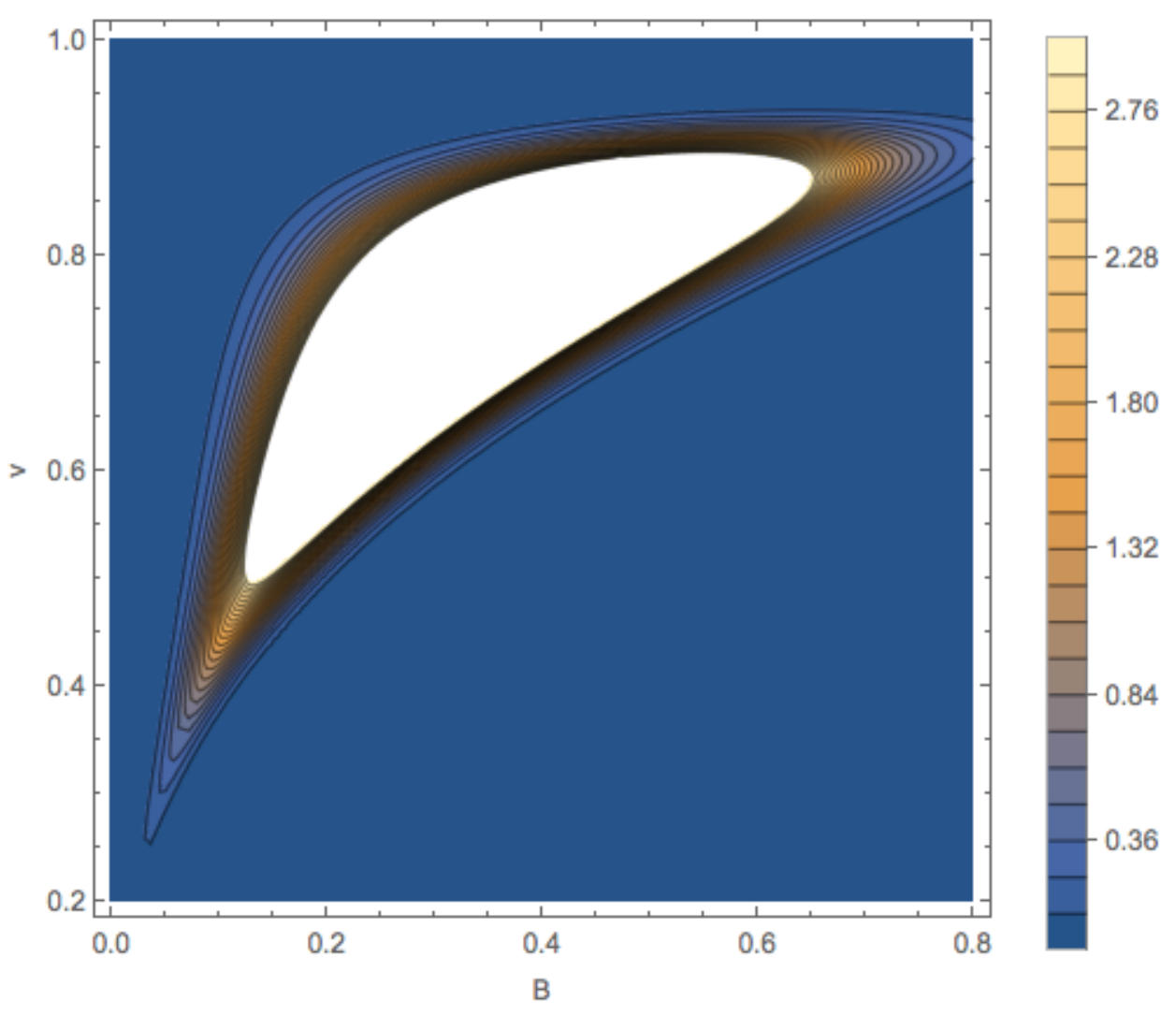}
\caption{Probability density function for a worldline centered at $(p,q)=(1,1)$ in the coordinates $(y,\eta)$ (left) and $(B,v)$ (right). In the plot $\sigma_p = \sigma_q=0.2$.}
\label{fig:prob_dens_yeta_11}
\label{fig:prob_dens_Bv_11}
\end{center}
\end{figure}

The bundle of worldlines resulting from the random extraction of $1000$ points according to the distribution  is depicted in figure \ref{fig:wordlines_bundle_11}. The probability density in the coordinates $(y,\eta)$ and $(B,v)$ are presented in figure \ref{fig:prob_dens_yeta_11}. As can be directly seen from these two figures the probabilistic behavior is different in these two set of coordinates, although the qualitative features of the probability densities are similar.

\subsection{Fuzzy worldlines in (2+1) dimensions and impact parameter}

Already from the (1+1) dimensional results discussed in the previous section we  start seeing that the distribution of fuzzy worldlines gets larger if the worldline is farther from the spatial origin at time $x^{0}=0$. In this section we are able to make this observation more precise by studying the distribution of fuzzy worldlines in $(2+1)$ dimensions and looking at the behaviour of the probability distribution of the impact parameter.

By using the relations \eqref{eq:vi}-\eqref{eq:Bi} and \eqref{eq:qpdefinition} one can write the impact parameter of two worldlines in terms of the coordinates $(q^{i},p^{i})$:
 In particular, in (2+1) dimensions the impact parameter  of a generic worldline $w$ characterized by coordinates $(q^{1},q^{2},p^{1},p^{2})$ with the worldline $w_0$, which has coordinates  $(q^{1},q^{2},p^{1},p^{2})=(0,0,0,0)$, equation \eqref{eq:distw0explicit}, reads:
\begin{equation}
\label{eq:distpq2+1}
\beta(w_0,w) = \frac{\left( \cosh p^1 - \frac{1}{\cosh p^2} \right)^{2} \left( q^2 \frac{\sinh p^1}{\cosh p^{2}} - q^1 \cosh p^1 \tanh p^2 \right)^2}{(\sinh p^1)^2 + (\tanh p^2)^2}\,.
\end{equation}

\begin{figure}[htbp]
\begin{center}
\includegraphics[width=0.32\textwidth]{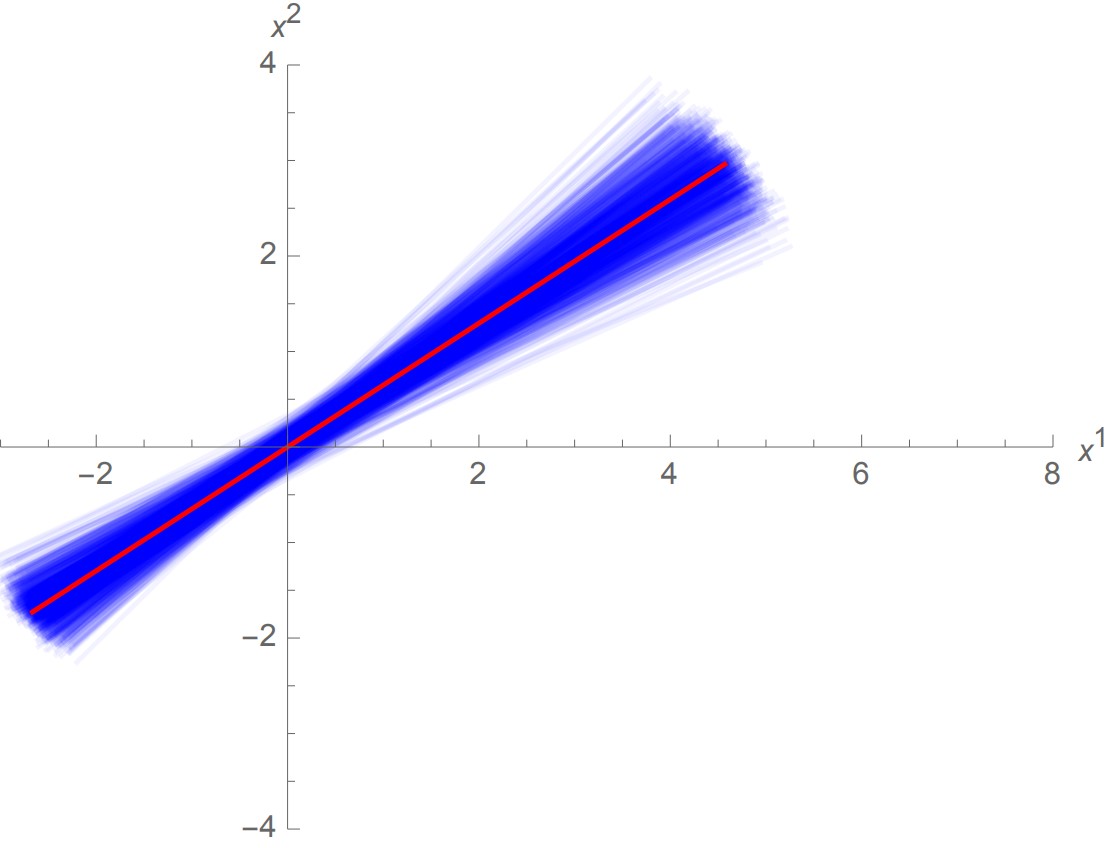}
\includegraphics[width=0.32\textwidth]{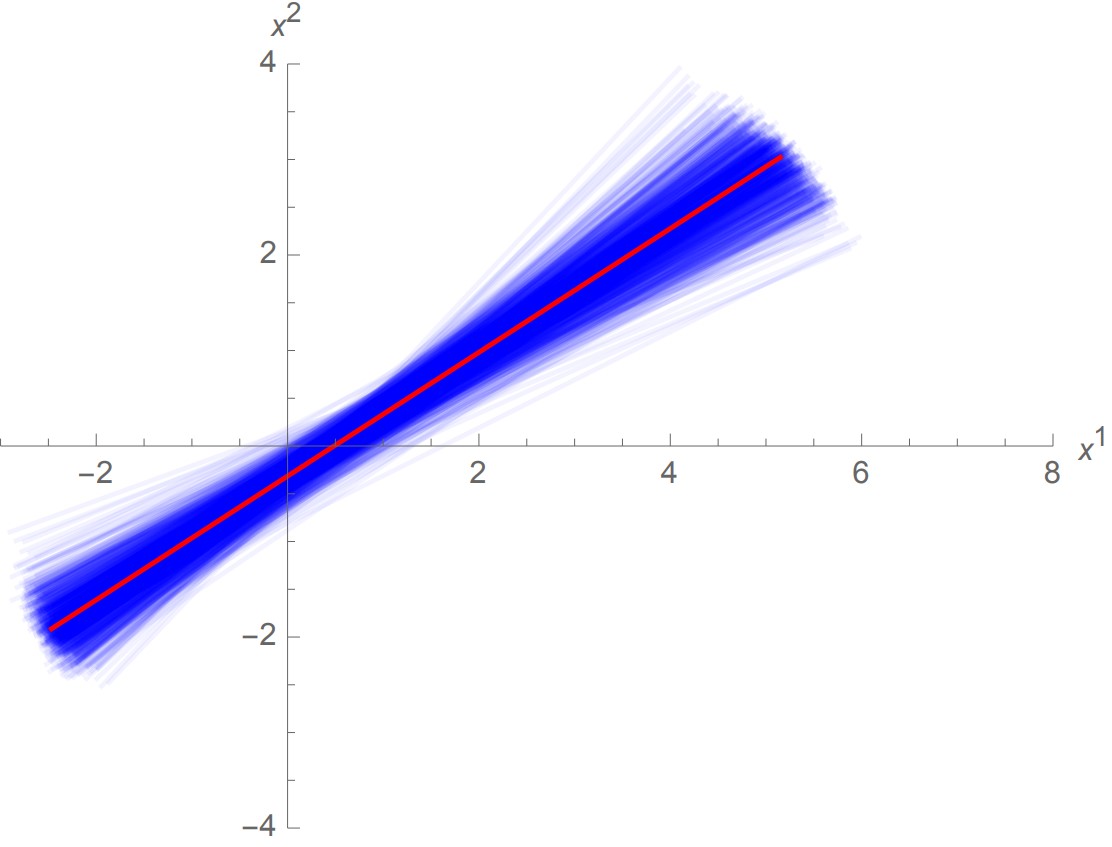}
\includegraphics[width=0.32\textwidth]{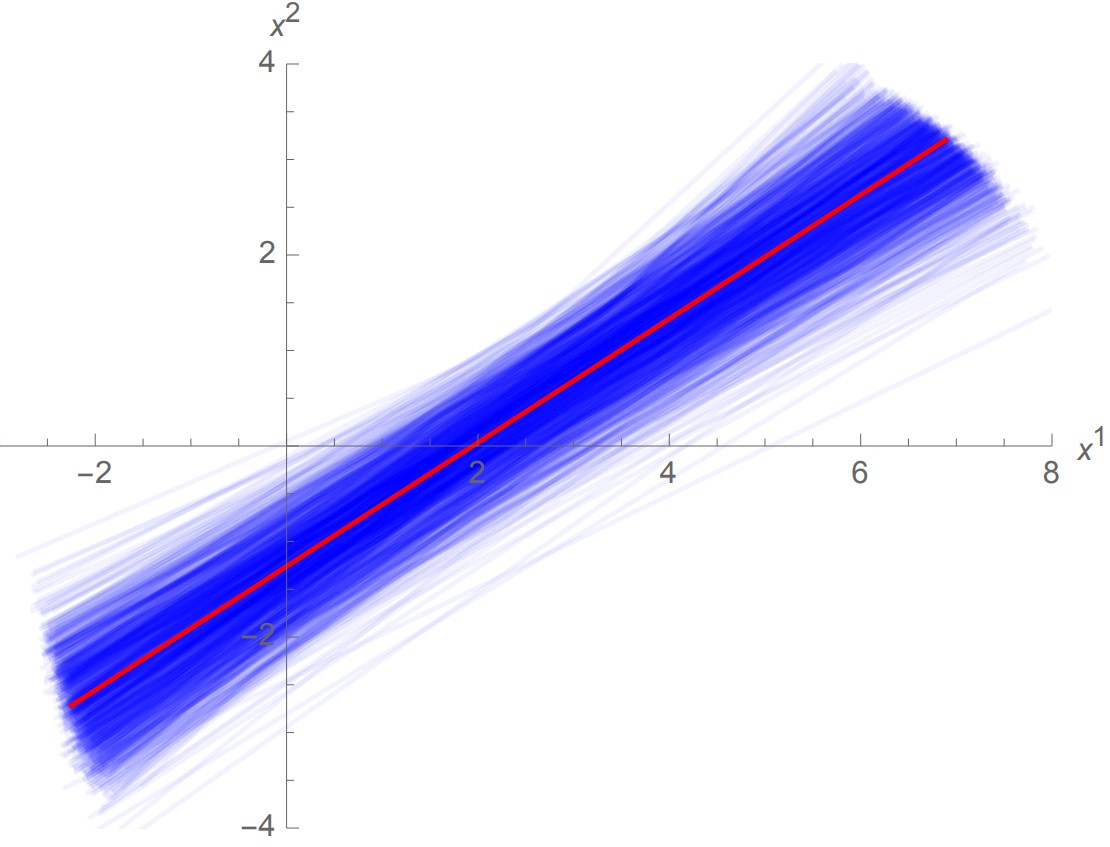}
\caption{Bundle of worldlines resulting from the random extraction of 1000 points from a Gaussian distribution of the form~(\ref{eq:WignerGaussian2D}), centred in $(p^{1},p^{2},q^{1},q^{2})$ with variance $\sigma = 0.01$. The plots show the projection of the worldlines in the plane $(x^{1},x^{2})$. From left to right, the distributions are centred in: $(p^{1},p^{2},q^{1},q^{2})=(1,1,0,0)$,  $(p^{1},p^{2},q^{1},q^{2})=(1,1,1,1)$,  $(p^{1},p^{2},q^{1},q^{2})=(1,1,4,4)$, corresponding to $(v^{1},v^{2},B^{1},B^{2})=(0.76,0.49,0,0)$, $(v^{1},v^{2},B^{1},B^{2})=(0.76,0.49,0.58,0.06)$, $(v^{1},v^{2},B^{1},B^{2})=(0.76,0.49,2.32,0.24)$ respectively. The red worldlines correspond to the center of the distributions. Each worldline has an opacity of $0.05$.}
\label{fig:bundles}
\end{center}
\end{figure}

We compare the properties of the probability distribution of this impact parameter for three different bundles of worldlines, depicted in figure \ref{fig:bundles} and corresponding to Gaussian distributions in $(q^{i},p^{i})$ corresponding to the 2D Wigner quasiprobability of eq.~\eqref{eq:WignerGaussian2D}. These bundles are centred in worldlines with the same velocity $v^{i}$ (equivalently, the same values of the coordinates $p^{i}$). What differs is the distance from the spatial origin, since each distribution is centred on different values of the coordinates $q^{i}$.   Already looking at  figure \ref{fig:bundles}, it is apparent that the farther the center of the  distribution is from the spatial origin, the larger the variance in the worldlines of the bundle.

\begin{figure}[htbp]
\begin{center}
\includegraphics[scale=0.45]{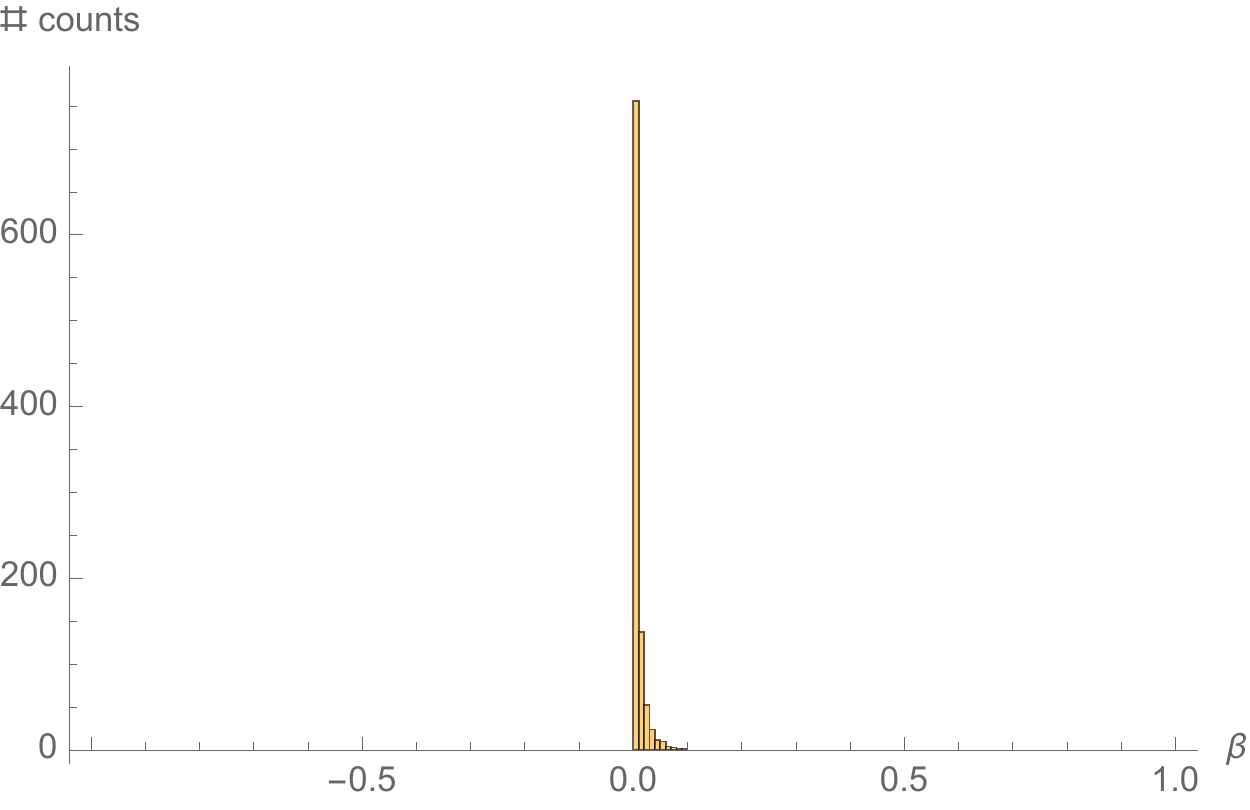}
\includegraphics[scale=0.45]{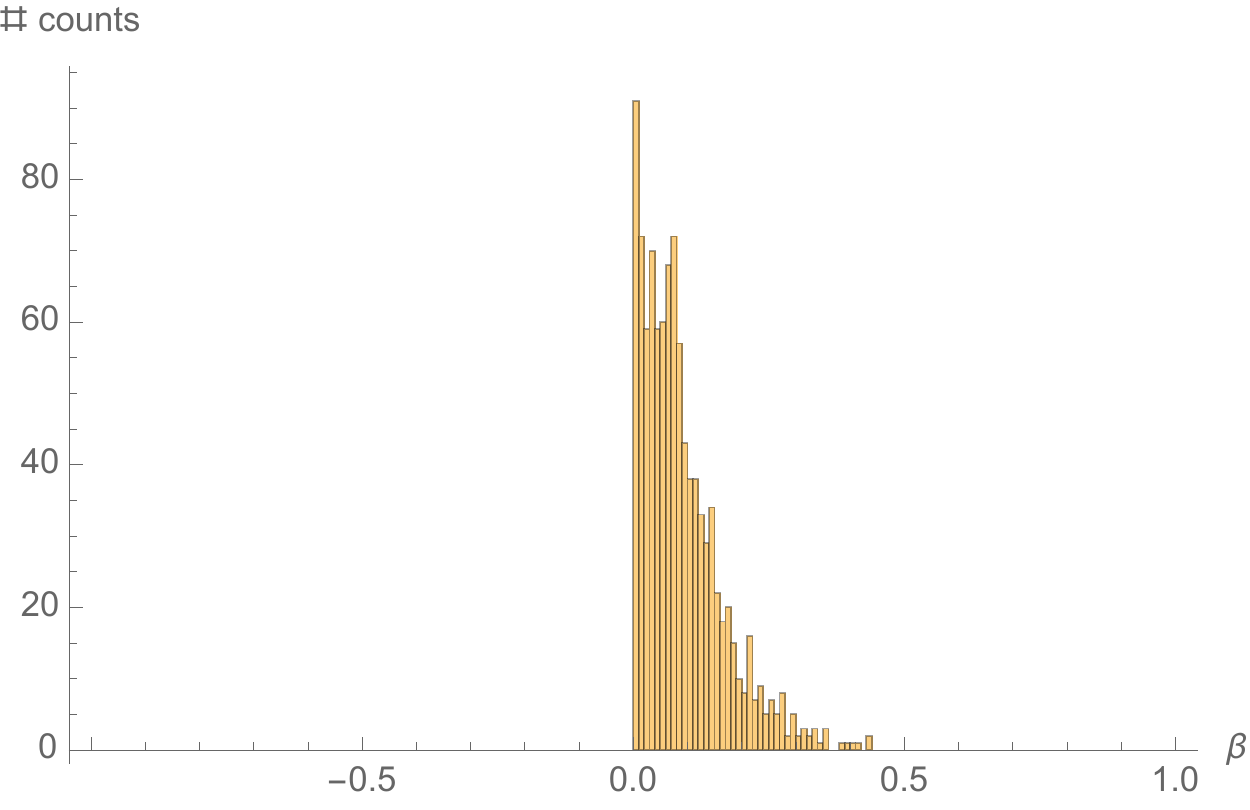}
\includegraphics[scale=0.45]{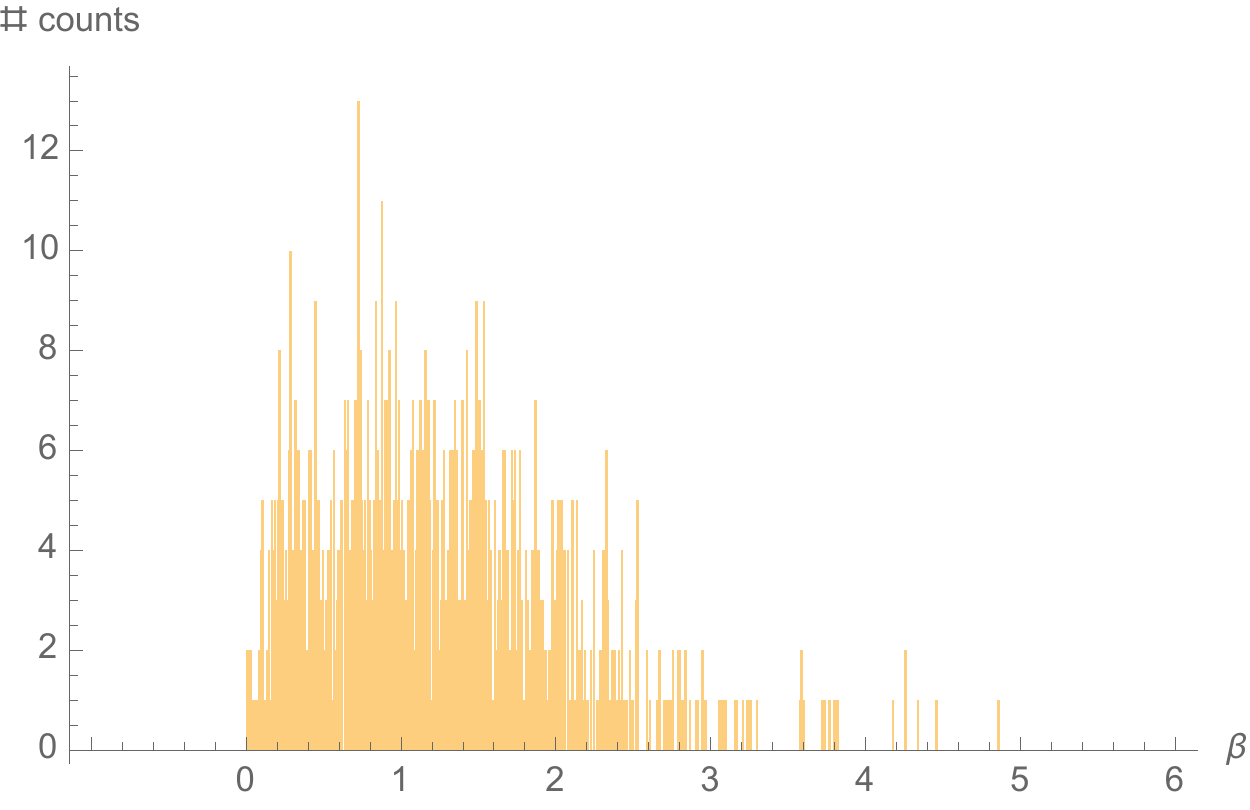}
\caption{Histogram of the values of the impact parameter for the bundle of worldlines resulting from the random extraction of 1000 points from a Gaussian distribution of the form~(\ref{eq:WignerGaussian2D}), centred in $(p^{1},p^{2},q^{1},q^{2})$ with variance $\sigma = 0.01$. From left to right, the distributions are centred in: $(p^{1},p^{2},q^{1},q^{2})=(1,1,0,0)$,  $(p^{1},p^{2},q^{1},q^{2})=(1,1,1,1)$,  $(p^{1},p^{2},q^{1},q^{2})=(1,1,4,4)$, corresponding to values of the impact parameter $\beta=0$,  $\beta=0.07$,  $\beta=1.12$, respectively. Note the different range of the horizontal axis of the bottom plot compared to the other two.
These histograms provide a first qualitative indication that the uncertainty $\delta \beta$ grows with the expectation value $\langle \beta \rangle$.}
\label{fig:impact}
\end{center}
\end{figure}

This observation is confirmed by the behaviour of the impact parameter $\beta$, whose histogram associated to the bundles of figure \ref{fig:bundles} is shown in figure \ref{fig:impact}. We see that the larger the classical value of the impact parameter, the larger  the variance of its distribution.

We interpret this effect as another instance of the well-known relative locality scenario \cite{AFKS2011deepening,AFKS2011principle,AAKRG2012relativelocality}, also discussed in \cite{LMMP2018localization} from the point of view of $\kappa$-Minkowski spacetime.~\footnote{{Strictly speaking, the relative locality effects have been studied in a non-quantum setting. This implies that it manifests itself as a systematic mismatch between the position of a far away event as inferred by some observer and its actual position with respect to the observer. In our quantum/fuzzy setting, relative locality  influences the uncertainty/fuzziness of  far-away events.}} In fact, we have already mentioned that we can define spacetime events by the crossing of two worldlines (say, $w_{1}$ and $w_{2}$), corresponding to $\beta(w_{1},w_{2})=0$. When the worldlines are fuzzy, as is the case in this section, the definition of the event becomes fuzzy as well: two worldlines that would classically cross might have tails of their fuzzy distributions which do not do so;  conversely, two worldlines which would classically not cross might have tails of their distributions which do. From the point of view of the impact parameter, this corresponds to saying that its distribution might show a nonzero probability for $\beta\neq0$ or for $\beta=0$, respectively.  Here we are  looking at the fuzzy event defined by the crossing of the sharp worldline $w_{0}$ (corresponding to an observer who is stationary in the spatial origin), with a generic fuzzy worldline $w$. Classically, one would have an event happening in the spatial origin only if $w$ crossed the spatial origin at some time. This is the case realized by the first bundle of worldlines, which indeed shows a corresponding distribution of $\beta$ that  peaks narrowly in $\beta=0$ (left-most panel of figure \ref{fig:impact}), thus leaving a small but nonzero probability of the event not taking place. In the other two panels, corresponding to fuzzy worldlines that pass increasingly farther from the spatial origin, the distribution of $\beta$ has an increasingly larger variance, allowing for a nonzero probability of having $\beta=0$ even in the situation depicted in the last panel of figures \ref{fig:bundles} and \ref{fig:impact}. That is, from the point of view of an observer at rest in the origin, the worldline of another particle appears the  `blurryer' the farther the particle passes from the observer, so much so that the observer might assign a nonzero probability to the worldline impacting on her detector even for far-away particles. In this sense, `locality' of the worldline $w$ becomes less and less precise the farther this worldline passes from the observer. 

In the next subsection we will study this issue quantitatively, by analyzing the statistical properties of the variable $\beta$.

\subsubsection{Quantitative description of the statistics of the impact parameter $\beta$}

\begin{figure}[htbp]
\begin{center}
\includegraphics[scale=0.62]{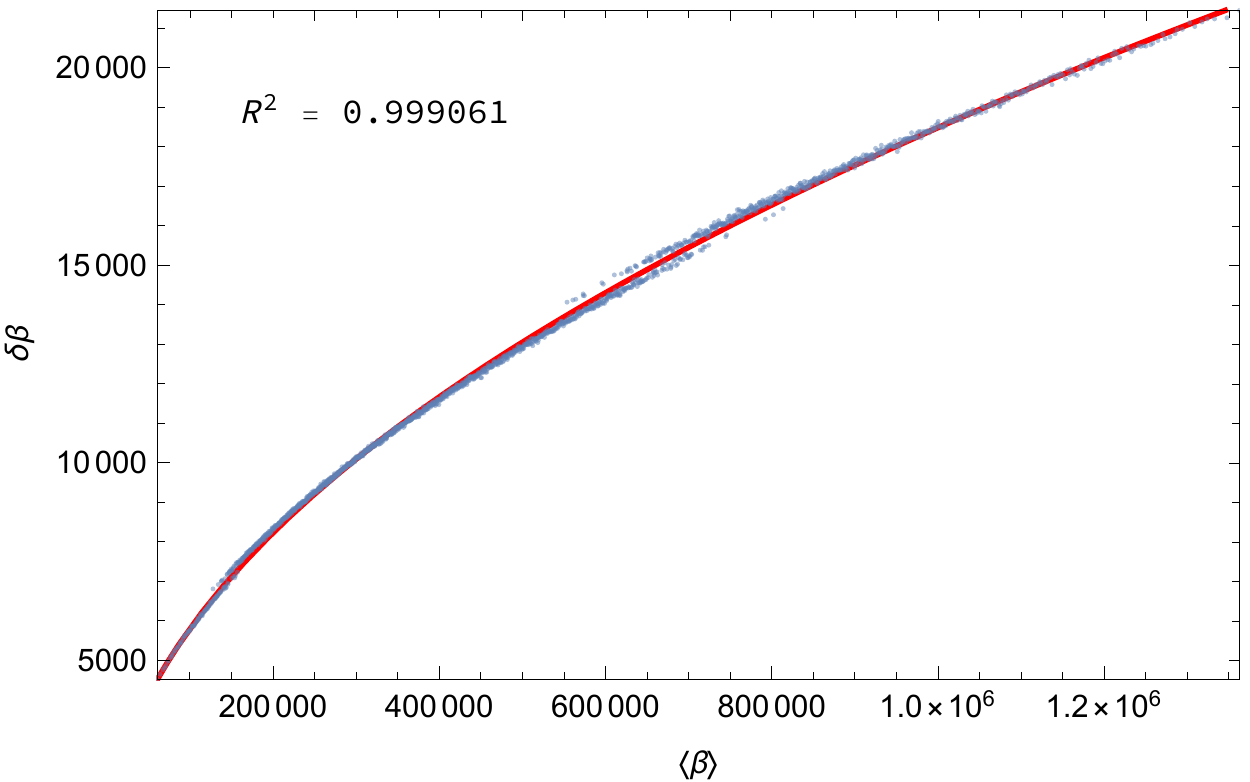}
\includegraphics[scale=0.62]{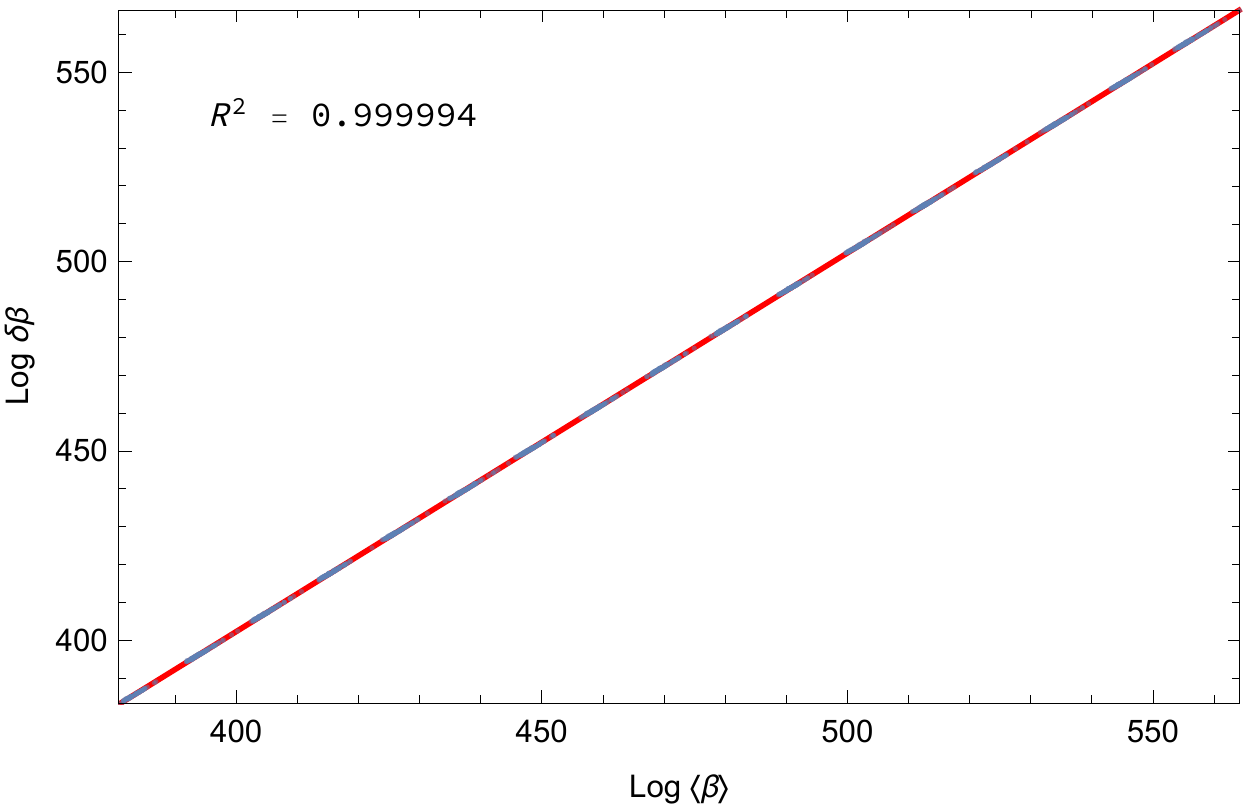}
\caption{Numerical calculation of the expectation value $\langle \beta \rangle$ and the variance $\delta \beta$. Left: taking $\sigma_q = 10$ and $\sigma_p = 10^{-5}$, with fixed $\bar p_1 = 0.856$ and $\bar p_2 = 1.3486$ and with $\bar q_i$ going from $-100 \sigma_q$ to $100 \sigma_q$. For each value of $\bar q_i$ a random sample of $50000$ values has been chosen.
Right: taking $\sigma_p = 2.7$ and $\sigma_q = 10^{-5}$, with fixed $\bar q_1 = 2.856$ and $\bar q_2 = 1.348$ and with $\bar p_i$ going from $-100 \sigma_p$ to $100 \sigma_p$. For each value of $\bar p_i$ a random sample of $50000$ values has been chosen. The scale is  logarithmic in both axes. The $R^2$ of the  model fit is included in both plots.
}
\label{fig:expectationvariance}
\end{center}
\end{figure}

The exact probability distribution of the impact parameter \eqref{eq:distpq2+1} when $(q^1,q^2,p^1,p^2)$ follow the Gaussian distribution corresponding to \eqref{eq:WignerGaussian2D} is rather complicated.   There are, however, two regimes in which the uncertainty $\delta \beta = \sqrt{\langle \beta^2 \rangle - \langle \beta \rangle^2}$ (the square root of the variance) has a  simple functional dependence on the expectation value $\langle  \beta \rangle$:
\begin{itemize}
\item When $\sigma_p \ll \sigma_q$, the uncertainty of $p^1$ and $p^2$ can be ignored, and they can be replaced, in the expression of $\beta$~(\ref{eq:distpq2+1}), with their expectation values $\bar p^1$ and $\bar p^2$:
$$
\beta = \frac{\left( \cosh \bar p^1 - \frac{1}{\cosh \bar p^2} \right)^{2} \left( q^2 \frac{\sinh \bar  p^1}{\cosh \bar p^{2}} - q^1 \cosh \bar  p^1 \tanh \bar  p^2 \right)^2}{(\sinh \bar p^1)^2 + (\tanh \bar p^2)^2}\,.
$$
Then $\beta$ is a quadratic form in $q_1$ and $q_2$, which are stochastic variables with a normal probability distribution. Therefore, the statistical properties of $\beta$  are expected to be those of a $\chi^2$ distribution, which satisfies:
\begin{equation}\label{eq:DeltaBetaSqrtMeanBeta}
\delta \beta \propto \sqrt{\langle\beta\rangle} \,.
\end{equation}
In other words, given a fixed value of the variances such that  $\sigma_p \ll \sigma_q$, and a fixed value for $\bar p^1$, $\bar p^2$, if we calculate $\langle \beta \rangle$ and $\delta \beta$ for different values of $\bar q_1$, $\bar q_2$, we will find that the pairs $(\langle \beta \rangle, \delta \beta)$ will sit near a square-root curve $\langle \beta \rangle \propto \sqrt{\delta \beta}$, where the proportionality factor depends on our choice of $\bar p^i$. This can be confirmed with numerical experiments, as shown in figure \ref{fig:expectationvariance}.

\item In the opposite regime $\sigma_p  \gg \sigma_q$, it is the variance of $q_1$ and $q_{2}$ that can be ignored, and they can be replaced with their expectation values in~(\ref{eq:distpq2+1}), namely:
$$
\beta = \frac{\left( \cosh p^1 - \frac{1}{\cosh p^2} \right)^{2} \left( \bar q^2 \frac{\sinh p^1}{\cosh p^{2}} - \bar q^1 \cosh p^1 \tanh p^2 \right)^2}{(\sinh p^1)^2 + (\tanh p^2)^2}\,.
$$
In this case $\beta$ has a complicated dependence on the normal variables $p^1$ and $p^2$, and we are not aware of an analytic argument for a particular dependence of $\delta \beta$ on $\beta$. However, numerically it turns out that, fixing the value of $\bar q_i$ and varying $\bar p^i$, the pairs $(\langle \beta \rangle, \delta \beta)$ end up very close to a straight line:
\begin{equation}\label{eq:DeltaBetaMeanBeta}
\delta \beta \propto \langle \beta \rangle \,,
\end{equation}
where the proportionality factor depends on $\bar q_i$, see figure \ref{fig:expectationvariance}.

\end{itemize}

The relationships we found between the variance and expectation values of $\beta$ in the different regimes have a counterpart in similar relationships that can be found when studying the set of states on the (1+1)-dimensional $\kappa$-Minkowski algebra, as was done in~\cite{LMMP2018localization}. 
 Among these states, the ones that saturate the uncertainty bound arising from the $\kappa$-Minkowski noncommutative spacetime~\eqref{eq:conmkappaMinkowski}, namely,
\begin{equation}
\delta \hat x^0 \delta \hat x^1 \geq \frac{1}{2 \kappa} \langle \hat x^1 \rangle \,,
\end{equation}
could be called squeezed states, according to how they distribute the uncertainty among $\hat x^0$ and $\hat x^1$. In this case, if we choose a set of states whose space and time uncertainties are proportional, \emph{i.e.} $\delta \hat x^0 \propto \delta  \hat x^1$, then
\begin{equation} \label{eq:DeltaX1SqrtMeanX1}
(\delta \hat x^1)^2 \propto  \langle \hat x^1 \rangle  ~~~ \Rightarrow ~~~ \delta \hat x^1  \propto \sqrt{ \langle \hat x^1 \rangle } \,,
\end{equation}
which is similar to the relation~(\ref{eq:DeltaBetaSqrtMeanBeta}). Note that (\ref{eq:DeltaX1SqrtMeanX1}) was found by assuming that $\delta \hat x^0 \propto \delta \hat x^1$, while (\ref{eq:DeltaBetaSqrtMeanBeta}) follows from $\sigma_p \ll \sigma_q$, and also after having fixed the expectation values of $\hat p^i$. The latter condition corresponds to considering a  worldline of a massive particle with fixed velocity (see relations \eqref{eq:vi} and \eqref{eq:qpdefinition}). It would be  interesting to investigate how these two sets of conditions are related.

On the other hand, one could fix the variance of $\hat x^0$ to a constant, and study the set of states with fixed $\delta \hat x^0$. Of course, for these states we have
\begin{equation} \label{eq:DeltaX1MeanX1}
\delta \hat x^1 \propto  \langle \hat x^1 \rangle \,,
\end{equation}
which is similar to~\eqref{eq:DeltaBetaMeanBeta}, which has been found by taking $\sigma_q \ll \sigma_p$, and fixing the expectation values of $\hat q_i$. Again, it is unclear how these conditions relate to the fixed variance we used to obtain~(\ref{eq:DeltaX1MeanX1}).

\section{Conclusions}

In this paper we made the first steps towards the development of a phenomenology of the $\kappa$-deformed `space of worldlines' first introduced in~\cite{BGH2019worldlinesplb}. Specifically, we introduced a noncommutative operator algebra associated to the Poisson bracket there defined as a direct byproduct of the $\kappa$-deformation. It is tempting to call this algebra a `quantization' of the homogeneous Poisson space of worldlines defined in~\cite{BGH2019worldlinesplb}, however one needs to be careful in recognizing that $\hbar$ does not play any role in our model, its role being played by $\kappa^{-1}$. So rather than `quantization' one should talk about a `noncommutativization', in the sense that our model is built by endowing the classical space of time-like worldlines associated to Minkowski spacetime with an additional Poisson structure that is controlled by the $\kappa^{-1}$ parameter. This is the novel ingredient which is assumed to encode the effects of the high-energy structure of spacetime, which vanish in the $\kappa\to\infty$ limit. The fuzziness effects are described in analogy with ordinary quantum mechanics by transforming this Poisson structure into a noncommutative algebra of observables for worldline coordinates in which $\kappa^{-1}$ plays the same role as  the Planck constant $\hbar$.

In this construction, covariance under the $\kappa$-Poincar\'e quantum group is guaranteed for both the spacetime and the space of worldlines. This implies that any two inertial observers are related by means of a quantum $\kappa$-Poincar\'e transformation. However, since by construction any observer is endowed with some uncertainty (in her position and/or velocity),  under a quantum Poincar\'e transformation her uncertainties will be affected by the group parameters of the quantum group inertial transformation, which are also $\kappa^{-1}$-noncommutative operators. 

Once our basic model was defined, we introduced a natural representation for the operator algebra, equipped with a Hilbert space and an inner product, which allow us to borrow the standard interpretational framework of quantum mechanics and associate expectation values and uncertainties of states to the various momenta of our noncommutative worldline operators.
We showed that our algebra admits perfectly localized states as improper eigenstates (i.e. limits of well-normalized wavefunctions), similarly to what happens in $\kappa$-Minkowski spacetime~\cite{LMMP2018localization,arXiv:1912.07098}. Such states correspond to worldlines of (massive) particles at rest. For worldlines corresponding to particles with a nonzero velocity one cannot define perfectly localized states, but squeezed states that minimize the uncertainties in  the phase space $(q,p)$ associated to the $\kappa^{-1}$-noncommutative Poisson structure for worldlines are possible. We then advocated the use of a semiclassical approach, in which Wigner quasiprobability distributions  on the phase space $(q,p)$  can be shown to associate Gaussian probability density functions to Gaussian wave packets. Then, even though the representation of our basic observables is rather complicated, we can extract the most salient statistical features of certain (Gaussian) states by studying the momenta of the corresponding classical probability density functions.

As an application, we studied the properties of the observable corresponding to the (2+1) dimensional ``impact parameter''  $ \beta$, which gives a measure of how close to each other two worldlines get, on Gaussian states. Classically, when this parameter is zero the two worldlines cross, thus defining a spacetime event. In the noncommutative setting, the variance of $\beta$ can be used to quantify the fuzziness of events. We highlight the existence of two regimes, in which the uncertainty in the observable $\beta$ goes like, respectively, the expectation value of $\beta$ or its square root. Which regime is realized depends on the state. In particular, it depends on whether the uncertainty on the $q$ variables dominates that on the $p$ variables or vice-versa. Interestingly enough, these results remind to those obtained in \cite{LMMP2018localization} in the context of $\kappa$-Minkowski spacetime, although in this paper only the (1+1)-dimensional case was studied. A rigorous understanding of the relation among these two approaches (the one presented in this paper based on observers and the one from in~\cite{LMMP2018localization}  focused on spacetime) would require an explicit description of the reconstruction of the spacetime (events) from the space of worldlines (inertial observers). 

From a somewhat different point of view, in order to study the effect of the $\kappa$-Poincar\'e quantum deformation on the causality of spacetime, the homogeneous space of light-like worldlines should be constructed and the implications for the propagation of light-rays analyzed. Moreover, the results here presented can be also generalized to the case of non-vanishing cosmological constant $\Lambda$ through the approach to $\kappa$-(A)dS groups and noncommutative spacetimes presented in~\cite{BHOS1994global,BHM2014plb,BHMN2017kappa3+1,BGH2019kappaAdS3+1}. Also, a similar approach could be considered for Newtonian and Carrollian spaces of worldlines through the contraction approach used in~\cite{SnyderG,ncNewCarr}. Work on these lines is in progress and will be presented elsewhere. 


\section*{Acknowledgements}

This work has been partially supported by Ministerio de Ciencia e Innovaci\'on (Spain) under grants MTM2016-79639-P (AEI/FEDER, UE) and PID2019 - 106802GB-I00 / AEI / 10.13039 / 501100011033, by Junta de Castilla y Le\'on (Spain) under grants BU229P18 and BU091G19, as well as by the Action CA18108 QG-MM from the European Cooperation in Science and Technology (COST). 


\end{document}